\documentclass[fleqn,usenatbib]{mnras}

\usepackage{newtxtext,newtxmath}

\usepackage[T1]{fontenc}

\DeclareRobustCommand{\VAN}[3]{#2}
\let\VANthebibliography\thebibliography
\def\thebibliography{\DeclareRobustCommand{\VAN}[3]{##3}\VANthebibliography}

\usepackage{graphicx}	
\usepackage{amsmath}	
\usepackage{pifont}
\usepackage{mathtools}
\usepackage{placeins}
\usepackage{float}
\usepackage{bm}
    
\usepackage{nccmath}

\usepackage{cleveref}
\crefname{figure}{Fig.}{Figs.}
\crefname{table}{Table}{Tables}

\title[Simulating the MRI on a moving mesh]{Simulating the magnetorotational instability on a moving mesh with the shearing box approximation}

\author[O. Zier and V. Springel]{%
Oliver Zier$^{1}$\thanks{E-mail: ozier@mpa-garching.mpg.de}
and Volker Springel$^{1}$
\\%
$^{1}$Max-Planck-Institut für Astrophysik, Karl-Schwarzschild-Straße 1, 85741 Garching, Germany\\
}

\date{Accepted XXX. Received YYY; in original form ZZZ}

\pubyear{2022}

\begin{document}
\label{firstpage}
\pagerange{\pageref{firstpage}--\pageref{lastpage}}
\maketitle

\begin{abstract}
The magnetorotational instability (MRI) is an important process in sufficiently ionized accretion disks, as it can create turbulence that acts as an effective viscosity, mediating angular momentum transport. Due to its local nature, it is often analyzed in the shearing box approximation with Eulerian methods, which otherwise would suffer from large advection errors in global disk simulations. In this work, we report on an extensive study that applies the quasi-Lagrangian,  moving-mesh code {\small AREPO}, combined with the Dedner cleaning scheme to control deviations from $\nabla\cdot\bm B=0$, to the problem of magnetized flows in shearing boxes. We find that we can resolve the analytical linear growth rate of the MRI with mean background magnetic field well.
In the zero net flux case, there is a threshold value for the strength of the divergence cleaning above which the turbulence eventually dies out, and in contrast to previous Eulerian simulations, the strength of the MRI does not decrease with increasing resolution. In boxes with larger vertical aspect ratio we find a mean-field dynamo, as well as an active shear current effect that can sustain MRI turbulence for at least 200 orbits. In stratified simulations, we obtain an active $\alpha\omega$ dynamo and the characteristic butterfly diagram. Our results compare well with previous results obtained with static grid codes such as {\small ATHENA}. We thus conclude that {\small AREPO} represents an attractive approach for global disk simulations due to its quasi-Lagrangian nature, and for shearing box simulations with large density variations due to its continuously adaptive resolution.
\end{abstract}


\section{Introduction}

The molecular viscosity of diffuse gas is by several orders of magnitude too small to explain the required amount of angular momentum transported in accretion disks. A possible solution is an effective viscosity that can be created by turbulence in the disk, and which in turn can be generated by different fluid instabilities \citep{shakura1973black,lynden1974evolution}. Besides the gravitational instability due to self-gravity \citep{gammie2001nonlinear}, there exist a plethora of other possible hydrodynamic and thermodynamic instabilities such as the Rossby wave instability \citep{Lovelace1999} or the convective overstability \citep{Klahr2014} that could be involved.

One of the most promising candidates for the main culprit is the magnetorotational instability (MRI) \citep{velikhov1959stability, chandrasekhar1960stability, fricke1969stability, balbus1991powerful} in ionized regions, which is a linear instability that is active in Keplerian-like shear flows and only requires a very small seed magnetic field to get started in the limit of ideal magnetohydrodynamics (MHD). The MRI's linear properties are nowadays well understood  \citep{balbus1991powerful, balbus1992powerful, curry1994global, goodman1994parasitic, kersale2004global} but its nonlinear behaviour is considerably more complicated and requires an analysis with computer simulations. Previous studies can be broadly categorized into global simulations that simulate the whole disk and into local simulations that compute only a small patch of the disk using the shearing sheet approximation \citep{hill1878collected,goldreich1965ii}. While the former allow capturing of global effects such as accretion, and the formation of winds and jets, they are also very expensive and only allow the analysis of a small part of the parameter space \citep{penna2010simulations, hawley2011assessing, hawley2013testing, parkin2013global,disco2016, deng2020global}.
Local simulations, in contrast, allow much higher resolution and are also cheaper, which means a much larger parameter space can be analyzed.

Shearing box simulations can again be split again into two groups: Stratified simulations that take the vertical gravitational force of the central object into account, and unstratified simulations neglecting this force component so that the focus lies on approximating the conditions in the mid-plane of the disk. An important property to characterize shearing box simulations is the mean magnetic flux, $\left< \bm B \right> = \int_V \bm B \, {\rm d}V$, whose radial and vertical component is conserved both for unstratified  and stratified simulations except if there are outflows in the vertical direction. The azimuthal component on the other hand is only conserved if there is no mean radial field. These properties motivate the definition of simulations with net-flux \citep[NF, $\left< \bm B \right> \neq 0$,][]{hawley1995local, sano2004angular, guan2009locality, simon2008simulations, gong2020}, and zero-net flux simulations \citep[ZNF, $\left< \bm B \right> = 0$,][]{hawley1996local, fromang2007mhd, simon2009viscous, bodo2011symmetries}. We note that this characterisation is however a simplification since in global simulations the mean net field in patches of the disk can change.

In the vertical NF case of unstratified simulations, one can first observe an exponential growth of so-called channel modes, which are a solution to the nonlinear ideal MHD equations. They grow until parasitic (secondary) instabilities destabilize them  \citep{goodman1994parasitic} and turbulence sets in. This turbulence alone decays but the background net field can revive the MRI which leads to  self-sustaining cycles. For the ZNF case, this background field is missing, and therefore perturbations in the velocity and magnetic field have to mutually sustain each other. Since the unstratified case is statistically symmetric it is harder to generate local mean fields, but  \citet{shi2016saturation} nevertheless found an active dynamo  in this situation, especially in tall boxes (large vertical aspect ratio). Three possible solutions are the stochastic $\alpha$-effect \citep{vishniac1997incoherent, silant2000magnetic, heinemann2011large},  the magnetic shear current effect \citep{rogachevskii2003electromotive, rogachevskii2004nonlinear, squire2015statistical} or the interplay of linear transient (nonmodal) growth of the MRI and the nonlinear transverse cascade that redistributes energy between different Fourier modes \citep{squire2014,Gogichaishvili2017,Gogichaishvili2018,Mamatsashvili2020,Held2022}.

In the stratified case, the system becomes anisotropic and consists typically of a turbulent, MRI unstable mid-plane surrounded by a magnetically dominated corona \citep{shi2009numerically, guan2011radially}. A mean-field dynamo is active in the mid-plane, where the sign of the mean field changes periodically, which leads to the typical `butterfly' diagram \citep{brandenburg1995dynamo, stone1996three, hirose2006vertical, gressel2010mean, davis2010sustained, simon2011resistivity}. Different effects such as the $\alpha \omega$ dynamo \citep{vishniac2001magnetic, subramanian2004nonlinear} and turbulent pumping \citep{gressel2010mean} have been invoked to explain this behaviour but also the effects already discussed for the unstratified case can have an influence on this evolution.

\cite{fromang2007mhd} found that for the unstratified ZNF case in small boxes the strength of the MRI decreases if the resolution is increased, and that for infinite resolution the MRI could even completely die out. However, convergence can be regained by explicitly accounting for viscosity and magnetic diffusivity \citep{fromang2007mhdTwo}, and there seems to exist a critical magnetic Prandtl number below which the ZNF MRI dies out. Interestingly, \cite{shi2016saturation} found convergence in larger boxes which they attributed to a large-scale mean field created by the shear current effect. 

In most studies of the MRI in shearing boxes, Eulerian methods were used that can benefit from the constrained transport method \citep{evans1988simulation} to preserve the condition $\nabla \cdot \bm B = 0$ up to machine precision. But they suffer from advection errors especially in global simulations with large bulk velocities, and they also cannot easily increase their resolution in the shearing box using adaptive mesh refinement (AMR). Langrangian methods on the other hand require some type of cleaning method to keep $\nabla \cdot \bm B$ errors small \citep[but see][for a recent implementation of constrained transport for the moving mesh method]{mocz2014constrained}, but they can achieve manifest Galilei invariance and easily allow a constant mass resolution. \cite{deng2019local} applied the particle-based MFM \citep{Hopkins2015} and SPH methods to the MRI and were able to show that MFM is able to accurately describe the linear growth of the MRI in unstratified boxes, and also sustains MRI turbulence for the unstratified NF case. In unstratified ZNF simulations, the turbulence died out however for SPH as well as MFM, and while MFM was able to reproduce the butterfly diagram for a stratified simulation with a time-varying net azimuthal field in the mid plane for around 50 to 70 orbits, the turbulence eventually still died out. Furthermore, in stratified simulations with SPH unphysically strong azimuthal fields grew. These findings underline the particular sensitivity of this problem to numerical errors of various kinds, making it challenging to obtain accurate and robust results.

\cite{wissing2021magnetorotational} performed a similar study with a modified version of standard SPH, the so-called geometrically-averaged density SPH (GDSPH) that can improve the accuracy of SPH in systems with large density gradients, as for example in the stratified simulations \citep{wadsley2017gasoline2, wissing2020smoothed}. Interestingly, they were able to sustain turbulence in the unstratified ZNF case, but did not find a large-scale mean-field dynamo in tall boxes. The authors attributed this to a missing shear current effect, unlike found in  \cite{shi2016saturation}. In stratified simulations, they however obtained an active $\alpha \omega$ dynamo and also reproduced the butterfly diagram for 200 orbits. These latter results compare quite well with previous Eulerian methods, and similar to them, they also show a dependence on the numerical Prandtl number. However, since GDSPH could not reproduce the  shear current effect, and because the geometric density averaging has been demonstrated to be problematic in cooling flows \citep{Springel2002}, it is unclear how universally applicable this variant of SPH is.

The moving mesh method \citep{springel2010pur,weinberger2020arepo} is a Lagrangian approach that tries to combine the advantages of a Galilei invariant Lagrangian method with the high accuracy of the finite volume method typically employed in Eulerian codes. This makes it especially interesting for global disk simulations but also for local simulations with large density gradients that can benefit from the code's high flexibility to continuously adapt cell sizes, and to increase and decrease the local resolution by splitting and merging individual computational cells.  \cite{pakmor2013simulations} has shown in a global simulation that the code can accurately capture the linear growth of the MRI, but only recently in \cite{zier2022simulating} we implemented the shearing box approximation in this code, and, importantly, we considerably reduced its residual numerical noise by means of higher-order flux integrations. The latter are particularly important in situations where cell shapes are constantly distorted at a high rate, such as in strongly shearing flow.

The goal of this paper is to analyse the performance of this improved method for simulating the MRI in different setups, and to compare the results with those obtained with Eulerian methods as well as Lagrangian methods described in \cite{deng2019local} and \cite{wissing2021magnetorotational}. We also put a special emphasis on the tensorial turbulent transport coefficients that allow us to understand the evolution of mean fields and therefore active dynamo processes in more detail. The detailed verification of the code accuracy we aim for here is clearly also a prerequisite for trusting predictions obtained with the code in planned future global disk simulations. 

This paper is structured as follows: In Section~\ref{sec:numericalMethods} we introduce the moving mesh method and especially the shearing box approximation as implemented in the {\small AREPO} code. We also analyze the linear growth of the MRI and introduce different quantities we will use subsequently to characterize the nonlinear, saturated state of the MRI as a function of the divergence cleaning strength as well as numerical resolution. In Section~\ref{sec:unstratifiedSimulations}, we discuss unstratified shearing box simulations. We analyze both the cases with a vertical NF as well as the case with ZNF in a standard and tall box, and show that in the latter situation a large-scale mean-field dynamo as in \cite{shi2016saturation} becomes active that is created by the shear current effect. In Section~\ref{sec:stratifiedSimulation}, we run four different stratified simulations and demonstrate a sustained and active MRI with turbulence in the mid-plane for 200 orbits, while above the mid-plane we find a magnetically dominated corona. Finally, in Section~\ref{sec:discussionSummary} we discuss and summarize our results.

\section{Numerical methods}

\label{sec:numericalMethods}
\subsection{The shearing box approximation}

The shearing box approximation is widely used in the study of the magnetorotational instability. It allows a higher spatial resolution in comparison to global disk simulations, and additionally delivers clearly defined boundary conditions which simplify the comparison of different studies. To implement the shearing box, we simulate a small patch of a disk centred at radius $r_0$. The rotational frequency of the disk at this point is given by $\Omega_0$ and we use a Cartesian coordinate system with $\hat{e}_x$ pointing in the radial direction, $\hat{e}_y$ in the azimuthal direction and $\hat{e}_z$ being perpendicular to the other two unit vectors. By transforming into the rotating system and linearizing the gravitational and centrifugal forces, the governing ideal MHD equations read as follows:
\begin{equation}
    \frac{\partial \bm U }{\partial t}+ \nabla \cdot \bm F(\bm U) = \bm S_{\rm grav} + \bm S_{\rm cor}.
    \label{eq:shearingBoxEquation}
\end{equation}
Here, we introduced the state vector $\bm U$, the flux vector $\bm F$, the source terms $ \bm S_{\rm grav}$ due to the gravitational and centrifugal forces, and a source term $\bm S_{\rm cor}$ describing the Coriolis force.
They are given by:
\begin{align}
\bm U  = \begin{pmatrix}
   \rho \\
   \rho \bm v  \\
   \rho e  \\
   \bm B\\
   \end{pmatrix},  \;\;\;\;\;\; 
   F(\bm U) =  \begin{pmatrix}
   \rho \bm v\\
   \rho \bm v \bm v^T + P -\bm B \bm B^T\\
   \rho e  \bm v + P \bm v -\bm B(\bm v \cdot \bm B)\\
   \bm B \bm v^T -\bm v \bm B^T,
   \end{pmatrix}, \\
\bm S_{\rm grav}  = \begin{pmatrix}
   0 \\
   \rho \Omega_0^2 \left(2q x \bm\hat{e}_x - z \bm\hat{e}_z\right)  \\
    \rho \Omega_0^2 \bm v \cdot \left(2q x \bm\hat{e}_x - z  \bm\hat{e}_z\right)   \\
   0\\
   \end{pmatrix}, \;\;\;\;\;\;  
   \bm S_{\rm cor} =  \begin{pmatrix}
   0\\
   -2 \rho \Omega_0  \bm\hat{e}_z \times \bm v\\
  0\\
  0,
   \end{pmatrix},
   \label{eq:shearingBoxSourceTerms}
\end{align}
where $\rho$, $\bm v$, $e$, $\bm B$, $P$ are the density, velocity, total energy per unit mass, magnetic field strength, and pressure, respectively.
The specific energy $e = u + \frac{1}{2} \bm v^2 + \frac{1}{2 \rho} \bm B^2$ consists of the thermal energy  per mass $u$, the kinetic energy density $\frac{1}{2} \bm v^2$, and the magnetic field energy density $\frac{1}{2 \rho} \bm B^2$. The pressure $P=p_{\rm gas} + \frac{1}{2}\bm B^2$ includes a thermal and a magnetic component. The system of equations is closed by the equation of state (EOS), which expresses $p_{\rm gas}$ as a function of the other thermodynamical quantities.

In this paper we use an isothermal EOS,
\begin{equation}
     p_{\rm gas} = \rho c_s^2,
\end{equation}
with constant isothermal sound speed $c_s = 1$. $\bm S_{\rm grav}$ depends on the shearing parameter
\begin{equation}
    q = - \frac{d \ln \Omega}{d \ln r},
\end{equation}
which simplifies to $q=3/2$ for the Keplerian case that we exclusively discuss in this paper. In general, we will use $\Omega_0 = 1$ and measure lengths in units of the scale height $H = c_s / \Omega_0$. $\bm S_{\rm grav}$ contains an optional term that represents a gravitational force in the $z$-direction. We will perform in this paper simulations with (stratified case) and without this term (unstratified case).

The above system allows for a ground-state solution with velocity field
\begin{equation}
    \bm v_0 = (0,-q\Omega_0 x,0),
    \label{eq:backroundShearFlow}
\end{equation}
at constant pressure and constant density field in an unstratified box, or with density field
\begin{equation}
    \rho (z) = \rho_0 \exp \left(- \frac{z^2}{2H^2}\right)
\end{equation}
in the case of a stratified box. To close the system of equations we also have to define boundary conditions (BCs).  In the $y$-direction, we use standard periodic BCs, and in the $z$-direction periodic BCs. In the $x$-direction we use the so-called shearing box BCs that are similar to standard periodic BCs but take into account the background shear flow (\ref{eq:backroundShearFlow}):
\begin{subequations}
\begin{equation}
    f(x,y,z,t) = f(x \pm L_x, y \mp w t, z,t),\;\;\;\; f\in \{\rho, \rho v_x, \rho v_z, \bm B\},
\end{equation}
\begin{equation}
    \rho v_y(x,y,z,t) = \rho v_y(x \pm L_x, y \mp w t, z,t) \mp \rho  w ,
\end{equation}
\begin{equation}
    e(x,y,z,t) = e(x \pm L_x, y \mp w t, z,t) \mp \rho v_y v_w + \frac{\rho w^2}{2} ,
\end{equation}
\label{eq:shearingBoxBoundaryConditions}
\end{subequations}
where $L_x$ is the box size in the $x$-direction and $w = q \Omega_0 L_x$. The boundary conditions therefore do not conserve the azimuthal momentum, nor the total energy or the azimuthal component of the volume-weighted averaged magnetic field \citep{gressel2007shearingbox}:
\begin{equation}
  \frac{\partial \left < B_y\right>}{\partial t}  = -\frac{w}{V}\int_{\partial x} B_x\, {\rm d}y \,{\rm d}z.
  \label{eq:evolutionBy}
\end{equation}
Here $\partial x$ denotes the boundary in the $x$-direction, and $V$ is the total volume of the box. Only in the case that the magnetic field has no mean radial component and $\nabla \cdot \bm B = 0$ holds, the azimuthal field is conserved. 

To solve equation (\ref{eq:shearingBoxEquation}) we use the moving mesh code {\small AREPO} \citep{springel2010pur, pakmor2016improving, weinberger2020arepo} that employs a moving, unstructured Voronoi mesh in combination with the finite volume method. We refer to \cite{zier2022simulating} for technical details of the implementation of the shearing box in this code. For all simulations, we use the  higher-order integration method for the flux as well as a second-order accurate Runge-Kutta time integration scheme recently introduced in the code \citep{zier2022simulating}.

\subsection{The divergence constraint of the magnetic field}
\label{subsec:divB}

A close inspection of equation (\ref{eq:shearingBoxEquation}) shows that if the initial conditions fulfill $\nabla \cdot \bm B =0$ this condition will remain true for all times. Numerical schemes that only find approximate solutions to the underlying analytical equations do not automatically fulfill this condition, and can sometimes be prone to developing numerical instability or large errors once a sizable divergence of the magnetic field appears. To reduce the influence of this error one can try to either remove it somehow, or to diffuse it away from its original site. {\small AREPO} supports both the Powell scheme \citep{powell1999solution, pakmor2013simulations} that diffuses the error, and the Dedner cleaning \citep{dedner2002hyperbolic, pakmor2011magnetohydrodynamics} approach that advects the error away and damps it. In contrast, the constrained transport method \citep{evans1988simulation, mocz2014constrained} avoids deviations from $\nabla \cdot \bm B =0$ to machine precision. Although this latter approach ensures negligible errors in the divergence constraint, it also tends to be somewhat more diffusive, and the construction of a constrained transport updating scheme algorithm is very difficult for meshes with changing topology.

The Powell scheme adds additional source terms to the underlying MHD equations that try to stabilize the system for the case of $\nabla \cdot \bm B \neq 0$:
\begin{equation}
    S_{\rm powell} = \begin{pmatrix} 0\\- \left(\nabla \cdot \bm B \right) \bm B \\ - \left(\nabla \cdot \bm B \right) (\bm v \cdot\bm B)\\ - \left(\nabla \cdot \bm B \right) \bm v\end{pmatrix}.
\end{equation}
The additional term in the induction equation depends on the absolute velocity $\bm v$, which unfortunately breaks the Galilei invariance of the moving-mesh method. This becomes problematic at the radial boundary of a shearing box set-up, since here the velocity of cells jumps discontinuously if they move through the boundary. Additionally, the source terms can modify and even generate a mean magnetic field in the vertical and radial directions due to this issue. This is especially problematic since according to equation~(\ref{eq:evolutionBy}) a mean radial field will continuously amplify the azimuthal component of the magnetic field.

The Dedner scheme adds an additional scalar field $\psi$ to the equations to be solved. The modified induction equation and the evolution of $\psi$ are given by:
\begin{equation}
    \frac{\partial}{\partial t} \begin{pmatrix}\bm B\\ \psi \end{pmatrix} + \nabla \begin{pmatrix}\bm B \bm v^T -\bm v \bm B^T +\psi \bm I \\ c_h^2 \bm B \end{pmatrix} = \begin{pmatrix}0\\ -c_h^2 / c_p^2 \psi \end{pmatrix}.
\end{equation}
Here $c_h$ is the velocity with which deviations from $\nabla \cdot \bm B = 0$ are diffused away, and $c_p$ defines the time scale over which $\psi$ decays. A larger value of $c_h$ typically leads to smaller errors in  $\nabla \cdot \bm B$ but also to higher numerical resistivity. Both effects can in principle influence the MRI. By default we set $c_h$ to the largest signal speed in our simulation, but we allow it to be multiplied with a prescribed constant factor $c_{h0}$ to analyze the effect of the cleaning speed on our results.

The signal speed is set equal to the velocity 
\begin{equation}
    c_f = \sqrt{c_s^2 + \frac{B^2}{\rho}}
    \label{eq:signalSpeed}
\end{equation}
of the fastest magneto-acoustic wave in the system. We use $c_p = \sqrt{2 c_h r}$, where $r$ is the effective radius of a cell. Since the Dedner scheme does not add any new source terms to the induction equation, the radial and vertical mean fields are conserved to machine precision. Although in the case of a magnetic field without a radial component the average azimuthal field should be conserved, this is not the case for $\nabla \cdot \bm B \neq 0$. A larger $c_h$ can in this case also help to reduce the magnitude of this spurious field component.

\subsection{Linear growth of channel flows}

The linear eigenmodes of the magnetorotational instability with a net vertical flux are called channel flows. They are solutions of the nonlinear ideal MHD equations, and their amplitude grows exponentially until parasitic instabilities destabilize them and a turbulent flow forms \citep{goodman1994parasitic}. We set up a box of size  $L_x = L_y = L_z =1$, initial background field $B= \bm (0,0,B_0)$, an isothermal equation of state with sound speed $c_s = 1$, and with a background shear flow (\ref{eq:backroundShearFlow}). We choose $\beta= 2 p_{\rm gas} /B_0^2 =84$ since in this case the wavelength $\lambda_{\rm fast} = 1$ of the fastest growing mode is equal to our box size, with a growth rate given by $0.75\,\Omega^{-1}$ \citep{latter2009mri}. As perturbation seed for the initial conditions we use:
\begin{subequations}
\begin{equation}
    \delta B = 0.001 \times B_0 \cos \left(2 \pi z \right) \frac{\hat{e}_x - \hat{e}_y}{\sqrt{2}},
\end{equation}
\begin{equation}
    \delta v = 0.001 \times \frac{3}{8 \pi}   \sin \left(2 \pi z \right) \frac{\hat{e}_x + \hat{e}_y}{\sqrt{2}}.
\end{equation}
\end{subequations}
As in \cite{deng2019local}, we run our simulation for $t= 8\, \Omega_0^{-1}$ and calculate the average growth rate $s_{\rm sim}$ in the simulation by using the amplitude of $\delta B$ at $t= 0$ and at $t= 8\, \Omega_0^{-1}$. We start with a Cartesian grid and rerun the simulation several times with different numbers of cells. As an error measure
we define  $e = (0.75 - s_{\rm sim}\Omega_0) /0.75$. 

\begin{figure}
    \centering
    \includegraphics[width=1\linewidth]{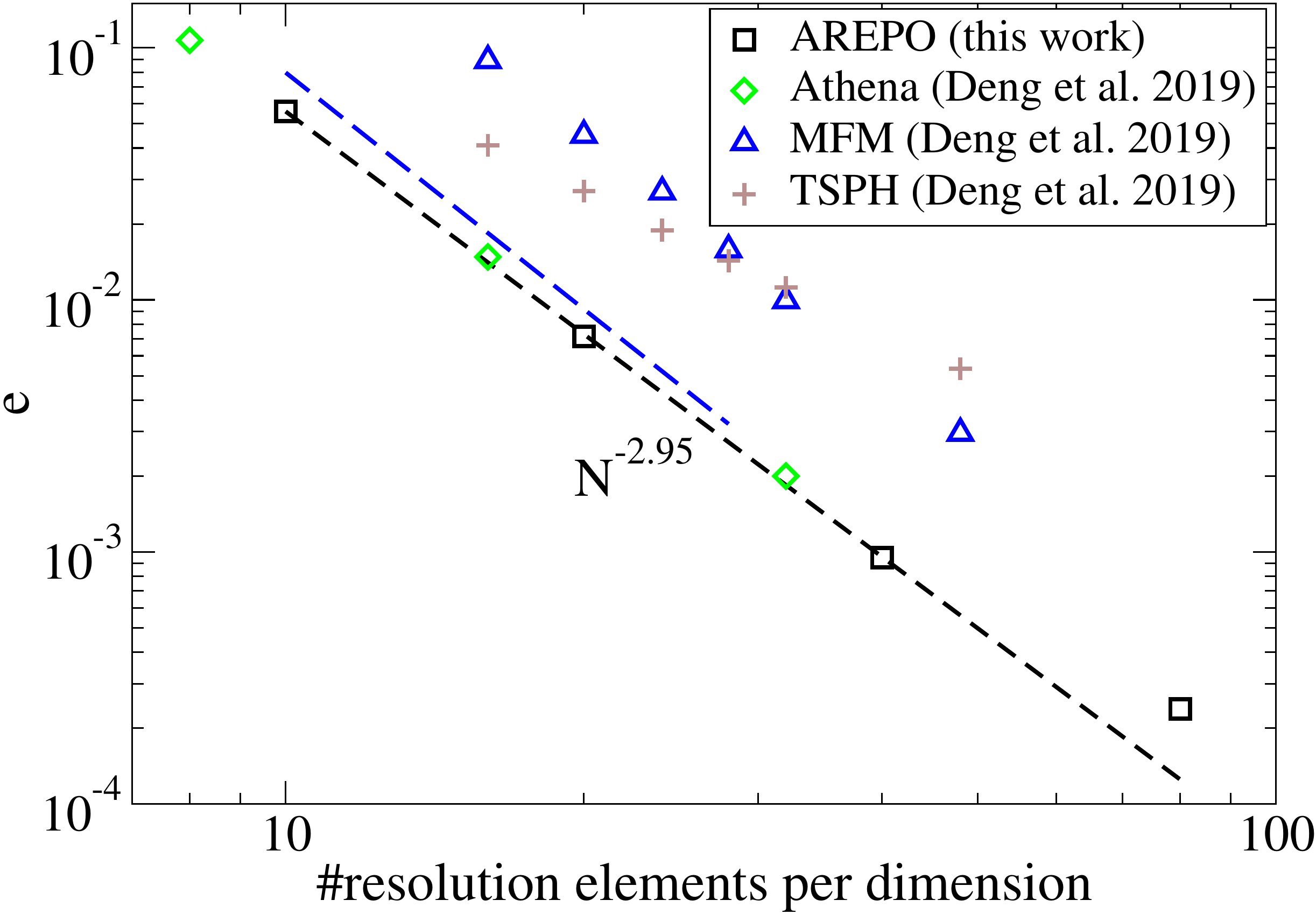}
    \caption{The error $e$ of the growth rate of the magnetic field of a channel flow as a function of the resolution.  The error converges with close to third order and the results compare well with results presented in \protect\cite{deng2019local} with the {\small ATHENA} code. We also show results with the MFM and TSPH implementations in {\small GIZMO} if we assume that each particle represents one resolution element. The blue-dashed line represents the results for MFM if we define the local resolution not as the mean particle distance but as the face-area weighted inter-neighbor separation \protect\cite[see][for details]{deng2019local}.}
    \label{fig:linearGrowthL1}
\end{figure}

In \cref{fig:linearGrowthL1} we show the error in the growth rate $s$ as a function of the employed resolution. Reassuringly, the results converge with close to third order to the analytical value. Also, the absolute values compare well with the results from the {\small ATHENA} code shown in \cite{deng2019local}. While the two particle-based methods SPH and MFM show larger absolute errors if we define the local resolution as the mean particle distance, MFM still manages to show the same third-order convergence as the grid-based methods.

To formally reduce the absolute error, \cite{deng2019local} defined the local spatial resolution in terms of the face-area weighted inter-neighbour separation, which leads to similar results as {\small ATHENA} and {\small AREPO}. We note, however, that this still implies a larger computational cost for the same spatial resolution, which becomes even worse than in standard MFM since the relatively large Wendland C4 kernel with 200 neighbours had to be used.

\begin{table*}
    \centering
    \begin{tabular}{c|c|c|c|c|c|c}
            \hline
        Type & Initial field & Box size & Base resolution & Res. multiplier& $c_{h0}$&Section \\
        \hline
        \hline
        Unstratified & $B_0 \hat{e}_z$, $\beta = 400 $ & $1\times 6.28\times 1$& $16 \times 100 \times 16$ & $1$, $2$, $3$, 4 &0.1, 0.5, 1, 2, 5&\ref{subsec:Unstraified_net_flux}\\
        Unstratified & $B_0 \hat{e}_z$, $\beta = 330 $ & $1\times 4\times 1$& $16 \times 64 \times 16$ & $1$,2,3,4 &0.1, 0.5, 1, 2, 5&App. \ref{app:netFluxMRIInSmallerBox}\\
        
        Unstratified & $B_0 \sin\left(2\pi x \right)\hat{e}_z$, $\beta = 400 $ & $1\times \pi\times 1$& $16 \times 50 \times 16$ & $1$, 2, 3, 4 &0.1, 0.5, 1, 2, 5&\ref{subsec:Unstraified_zero_net_flux}\\
        \hline
                Unstratified & $B_0 \sin\left(2\pi x \right)\hat{e}_z$, $\beta = 400 $ & $1\times 4\times 4$& $16 \times 64 \times 64$ & $1$, 2, 3 &0.1, 0.5, 1, 2, 5&\ref{subsec:Unstraified_zero_net_flux_tall}\\
                        \hline
                Stratified & $B_0 \hat{e}_y$, $\beta = 25 $ & $\sqrt{2}\times 4\sqrt{2} \times 24$& $\approx 1.5 \times 10^6$ cells& 1  & 0.5, 1&\ref{sec:stratifiedSimulation}\\
                Stratified & $B_0 \hat{e}_y$, $\beta = 25 $ & $\sqrt{2}\times 4\sqrt{2} \times 24$& $\approx 3 \times 10^6$ cells& 1 &0.5, 1&\ref{sec:stratifiedSimulation}\\
                        \hline
    \end{tabular}
    \caption{Overview of all simulations discussed in this paper. 
    The initial field strength is determined by the plasma beta $\beta= 2p_{\rm gas} /B_0^2$. For some simulations we analyze the influence of resolution by multiplying the amount of cells of the base resolution with a constant factor in all dimensions. We also analyze the effect of the cleaning speed $c_{h0}$ of the Dedner scheme on our results. In the stratified simulations we allow cells to merge and be split, and we enforce an approximately constant mass per cell.}
    \label{tab:overviewSimulations}
\end{table*}

\subsection{Analysis and overview of simulations}

To analyze our simulations we define the volume-weighted average of a quantity $X$ as
\begin{equation}
\left <X \right > = \frac{\int X  {\rm d}V}{\int  {\rm d}V},   
\end{equation}
as well as the temporal average of $X$,
\begin{equation}
\left <X \right >_t = \frac{\int X  {\rm d}t}{\int  {\rm d}t}.
\end{equation}
For the first quantity, we integrate over the whole simulation box if not stated otherwise, while for the second one we typically only integrate over the time interval during which the MRI is saturated in the nonlinear regime. For clarity, we will always mention the start of this averaging time interval.

To measure the angular momentum  transport and the saturation level of the MRI it is useful to calculate the Maxwell stress
\begin{equation}
\alpha_{M} = - \frac{B_x B_y }{ P},
\end{equation}
as well as the Reynolds stress
\begin{equation}
\alpha_{R} =  \frac{\rho v_x \delta v_y }{ P},
\end{equation}
where $P$ is the pressure and $\delta v_{y} = v_y - v_{y,0}$ is equal to the velocity relative to the background shear flow. 
A related quantity is the normalized magnetic stress:
\begin{equation}
\alpha_{\rm mag} = - \frac{\left <B_x B_y \right >}{ \left < B^2 \right >}.
\label{eq:defAlphaMag}
\end{equation}
As in \cite{shi2016saturation} and \cite{wissing2021magnetorotational}, we decompose the magnetic field into a mean field $ \overline{\bm B}$ and a turbulent field $\bm b$, where the first component is defined as  the horizontal average
\begin{equation}
    \overline{X} = \frac{\int X \, {\rm d}x  {\rm d}y}{\int  {\rm d}x  {\rm d}y}.
\end{equation}
While the volume integrals can be directly calculated using a Voronoi mesh, the horizontal average is more complicated. We address this by first binning our simulation data to a uniform Cartesian grid with typically twice the number of cells per dimension as in the initial conditions, followed by carrying out the integral using this mesh.

\begin{figure*}
    \centering
    \includegraphics[width=0.8\linewidth]{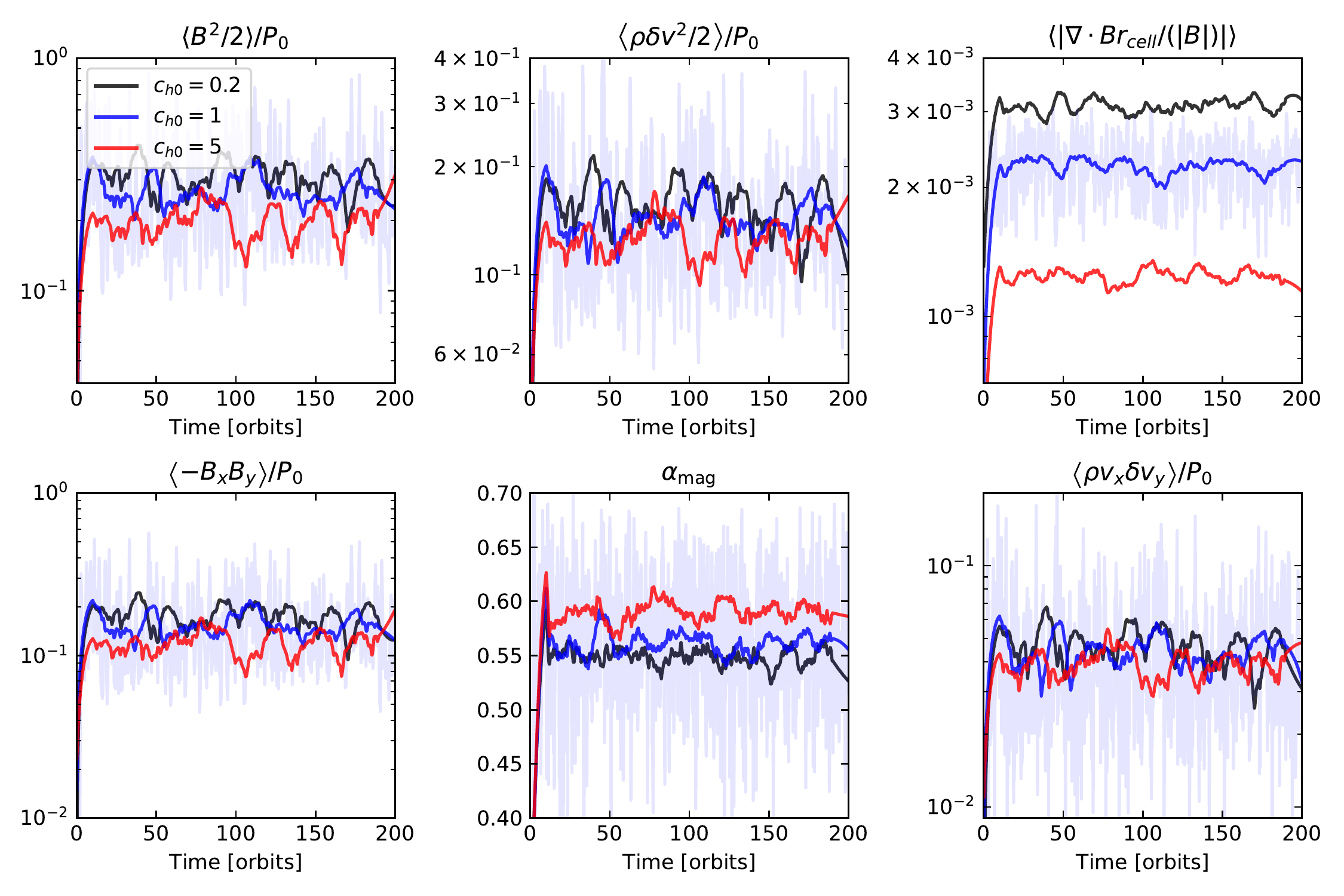}
    \caption{The temporal evolution of several volume weighted quantities for unstratified simulations with a net vertical magnetic field, box size $1\times 6.28 \times 1$ and $48$ cells per scale height.  We vary the strength of the Dedner cleaning $c_{h0}$, as labelled. The shown quantities are (from left to right, and top to bottom): Magnetic field energy density, kinetic energy density, relative $\nabla \cdot \bm B$ error, Maxwell stress, normalized Maxwell stress (\ref{eq:defAlphaMag}) and Reynolds stress. We have smoothed the curves over 20 orbits using a Savitzky–Golay filter, and show the original unsmoothed curve for one example case as a transparent line in the background.}
    \label{fig:evolution_net_field_2pi}
\end{figure*}

\begin{figure*}
    \centering
    \includegraphics[width=1\linewidth]{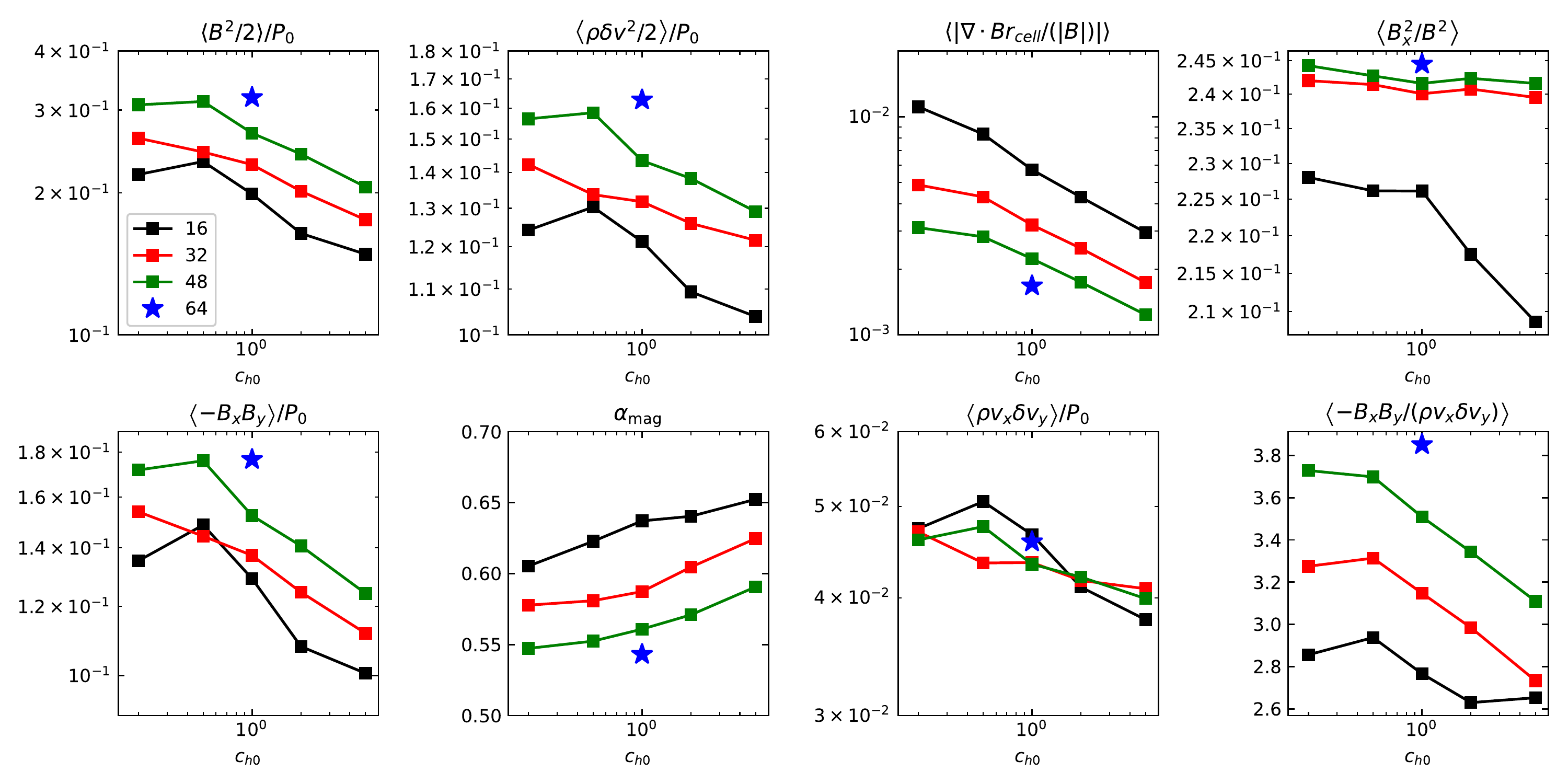}
    \caption{The temporal average of different quantities for unstratified simulations with a net vertical magnetic field and box size $1\times 6.28 \times 1$, as a function of the strength $c_{h0}$ of the Dedner cleaning. All quantities are averaged over 150 orbits starting after 50 orbits. We also vary the resolution with 16, 32, 48 and 64 cells per scale height, as labelled. For the highest resolution we only performed one simulation with $c_{h0} = 1$ due to the high computational costs. Besides the quantities already shown in \cref{fig:evolution_net_field_2pi}, we also include the  ratio between radial and total magnetic field energy (top right) and the ratio of the Maxwell and Reynolds stress (bottom right).}
    \label{fig:overview_avg_NetFlux_2Pi}
\end{figure*}

As we have discussed in Section~\ref{subsec:divB}, our implementation of the MHD equations does not preserve the condition $\nabla \cdot \bm B = 0$. To measure the potential impact of numerically induced magnetic monopole errors we define the relative divergence error
\begin{equation}
    \epsilon_{\nabla \cdot B,i} = \frac{\left(\nabla \cdot \bm B_i\right) r_i}{\left |\bm B_i \right|},
\end{equation}
where $r_i = \left[3V_i /(4 \pi)\right]^{1/3}$ is the effective radius of the Voronoi cell $i$.
In the case of $\left< B_x\right> = 0$ the mean azimuthal component  $\left< B_y\right>$ of the magnetic field can only change for $\nabla \cdot \bm B \neq 0$.
The evolution of  $\left< B_y\right>$ can therefore be used to measure the impact of divergence errors on our simulation results.

Linear stability analysis leads to the definition of the quality factor \citep{noble2010dependence}
\begin{equation}
    Q_i= \frac{\lambda_{\rm MRI}}{h} = \frac{2 \pi v_{A,i}}{\Omega h},
\end{equation}
where  $\lambda_{\rm MRI}$ is the characteristic wavelength, $v_{A,i}$ is the $i$-component of the Alfven velocity, and $h$ is the local spatial resolution. Similar to static grid codes, where $h$ is typically set to the grid cell size \citep{hawley2011assessing, parkin2013equilibrium}, we define it here as the effective diameter $d= 2\left(\frac{V}{4/3 \pi}\right)^{1/3}$ of each Voronoi cell. Although  $Q > 6$ is sufficient to properly resolve the linear growth of the MRI (see also the previous section), $Q_z > 10$ and $Q_y > 20$ are required to achieve convergence in the stresses for the case of a stratified net flux simulation \citep{hawley2011assessing}.

Although the definition of $Q_i$ based on linear theory and the net flux case makes its application to the nonlinear regime questionable (especially for the zero net flux case),  it is still a useful indicator to estimate whether the MRI can still be resolved. This applies especially in the stratified case, where the density and therefore spatial resolution strongly varies within the simulation box. In Table~\ref{tab:overviewSimulations} we give an overview of all the primary simulations performed for this paper, together with their principal numerical parameters.

\subsubsection{Tensorial transport coefficients}

To better understand the influence of the small-scale fluctuations on the large-scale field it is useful to use the concept of mean-field theory \citep{moffatt1978field, parker2019cosmical, krause2016mean, ruzmaikin1988magnetism, brandenburg2005astrophysical}.
In the following, we will mostly follow the discussion in \cite{wissing2021magnetorotational} to which we refer for a more in-depth coverage. By averaging the induction equation, the evolution of the mean magnetic field is given by:
\begin{equation}
\label{eq:meanfieldinduc}
\frac{\partial\overline{\bm B}}{\partial t}=\nabla \times [(\bm v-\bm v_0) \times \overline{\bm B}] + \nabla \times \mathcal{\bm E}.
\end{equation}
Here $\mathcal{\bm E}$ is the electromotive force (EMF) generated by the fluctuations in the velocity and magnetic field:
\begin{equation}
 \mathcal{\bm E}=\overline{ (\bm v - \bm v_0) \times\bm b}.
\end{equation}
By splitting the velocity and magnetic field fluctuations into  components independent of the mean-field and components linearly dependent on the applied mean-field, we can Taylor expand this expression to leading order under the assumptions of scale separation and the absence of correlations between the independent components of the mean magnetic field and the velocity perturbations:
\begin{equation}
 \mathcal{E}_i = \alpha_{ij} \overline{B_j} - \eta_{ij} \overline{J_j} + ....
\label{eq:EMFFApproximation}
\end{equation}
Here we introduced the mean-field current density
\begin{equation}
\overline{\bm J} = \overline{\nabla \times \bm B}
\end{equation}
and the tensorial transport coefficients $\alpha$ and $\eta$. Since $\overline{B_z} = \overline{J_z} = 0$, this simplifies to:
\begin{equation}
\label{eq:emfx}
\mathcal{E}_x=\alpha_{xx}\overline{B_x}+\alpha_{xy}\overline{B_y}-\eta_{xx}\overline{J_x}-\eta_{xy}\overline{J_y},
\end{equation}
\begin{equation}
\label{eq:emfy}
\mathcal{E}_y=\alpha_{yx}\overline{B_x} +\alpha_{yy} \overline{B_y}-\eta_{yx}\overline{J_x} -\eta_{yy} \overline{J_y}.
\end{equation}
By multiplying the two equations with $\left\{\overline{B_x},\, \overline{B_y},\, \overline{J_x},\, \overline{J_y}\right\}$ we obtain 8 equations for the 8 components of $\alpha$ and $\eta$ which are in general functions of height $z$ and time $t$.

A direct solution of the system of linear equations leads to quite noisy measurement results, which can be improved by using the approximations $\alpha_{xx} = \alpha_{yy}$, $\eta_{xx} = \eta_{yy}$ \citep[which can be justified as in][]{hubbard2009alpha, gressel2010mean} and $\alpha_{yx} = 0 = \eta_{xy}$, which is justified due to $\overline{B_x} \ll \overline{B_y}$ \citep{squire2015generation}. For the unstratified simulations, we then determine averaged transport coefficients in the $z$-direction by integrating the linear system of equations over the whole box in the $z$-direction, and assuming $z$-independent coefficients. This leads to a single linear system of equations for each time step, and thus gives access to the temporal evolution of the transport coefficients. With this in hand, we perform a standard temporal average of the transport coefficients, which is equivalent to averaging over many different realizations of the turbulent state \citep{squire2015generation}.

For stratified simulations we allow nonzero $\alpha_{yx}$, $\eta_{xy}$ and $\alpha_{xx} \neq \alpha_{yy}$, as in \citet{wissing2021magnetorotational} in order to simplify a direct comparison with other studies. In this case, we assume the transport coefficients to be independent of time, and try to calculate their structure as a function of $z$. For each $z$-value, we solve the overdetermined system of $8\times N$ equations for the 7 independent transport coefficients by minimizing the residual, where $N$ is the number of snapshots we use for our calculation. In general, we typically have five snapshots per orbit.

By inserting equation~(\ref{eq:EMFFApproximation}) into (\ref{eq:meanfieldinduc}), the evolution of the averaged quantities is given by:
\begin{equation}
\label{eq:dBdtx}
\frac{\partial\overline{B_x}}{\partial t}=-\partial_z(\alpha_{yx}\overline{B_x})-\partial_z(\alpha_{yy}\overline{B_y})+\partial_z(\eta_{yx}\overline{J_x})+\partial_z(\eta_{yy}\overline{J_y}),
\end{equation}
\begin{equation}
\label{eq:dBdty}
\frac{\partial\overline{B_y}}{\partial t}=-q \Omega \overline{B_x}+\partial_z(\alpha_{xx}\overline{B_x})+\partial_z(\alpha_{xy}\overline{B_y})-\partial_z(\eta_{xx}\overline{J_x})-\partial_z(\eta_{xy}\overline{J_y}).
\end{equation}
The components $\alpha_{xx}$ and $\alpha_{yy}$ are the main drivers of the $\alpha$ effect that can lead to the well-known $\alpha\omega$ dynamo  in combination with differential rotation. It requires a statistical symmetry breaking, e.g. a stratification or a net helicity \citep{pouquet1976strong, moffatt1978field, brandenburg2005astrophysical}. We, therefore, expect them to be zero in our unstratified simulations and antisymmetric relative to the mid-plane in the stratified simulations. The antisymmetric components $\alpha_{xy}$ and $\alpha_{yx}$ define the diamagnetic pumping term
\begin{equation}
    \gamma_z = \frac{1}{2} \left(\alpha_{yx} - \alpha_{xy}\right)
    \label{eq:diamagneticPumpingTerm}
\end{equation}
that describes the transport of the mean fields by the turbulent components. It is expected to be non-zero for the stratified case. The diagonal terms $\eta_{xx}$, and $\eta_{yy}$ describe the diffusion of the mean field, while the off-diagonal coefficients $\eta_{xy}$ and $\eta_{yx}$ are responsible for the dynamo produced by the $\Omega \times J$ effect \citep{radler1969elektrodynamik} and the shear current effect \citep{rogachevskii2003electromotive, squire2015statistical, squire2015generation, squire2015electromotive}. The latter requires $\eta_{yx} <0$.

\section{Unstratified simulations}
\label{sec:unstratifiedSimulations}

In this section we discuss simulations without the gravitational term in the vertical direction (see equation~\ref{eq:shearingBoxSourceTerms}). We start with a uniform Cartesian grid, set the initial velocity field to the ground state of the shearing box (\ref{eq:backroundShearFlow}), and use an initially uniform density $\rho = 1$ with sound speed $c_s = 1$. To seed the MRI, we add random noise of maximum amplitude $0.05\, c_s$ to the background shear flow of every cell. We characterize the strength of the initial magnetic field with the volume-averaged plasma beta $\beta = 2 p_{\rm gas} /B^2$.

\begin{figure}
    \centering
    \includegraphics[width=1\linewidth]{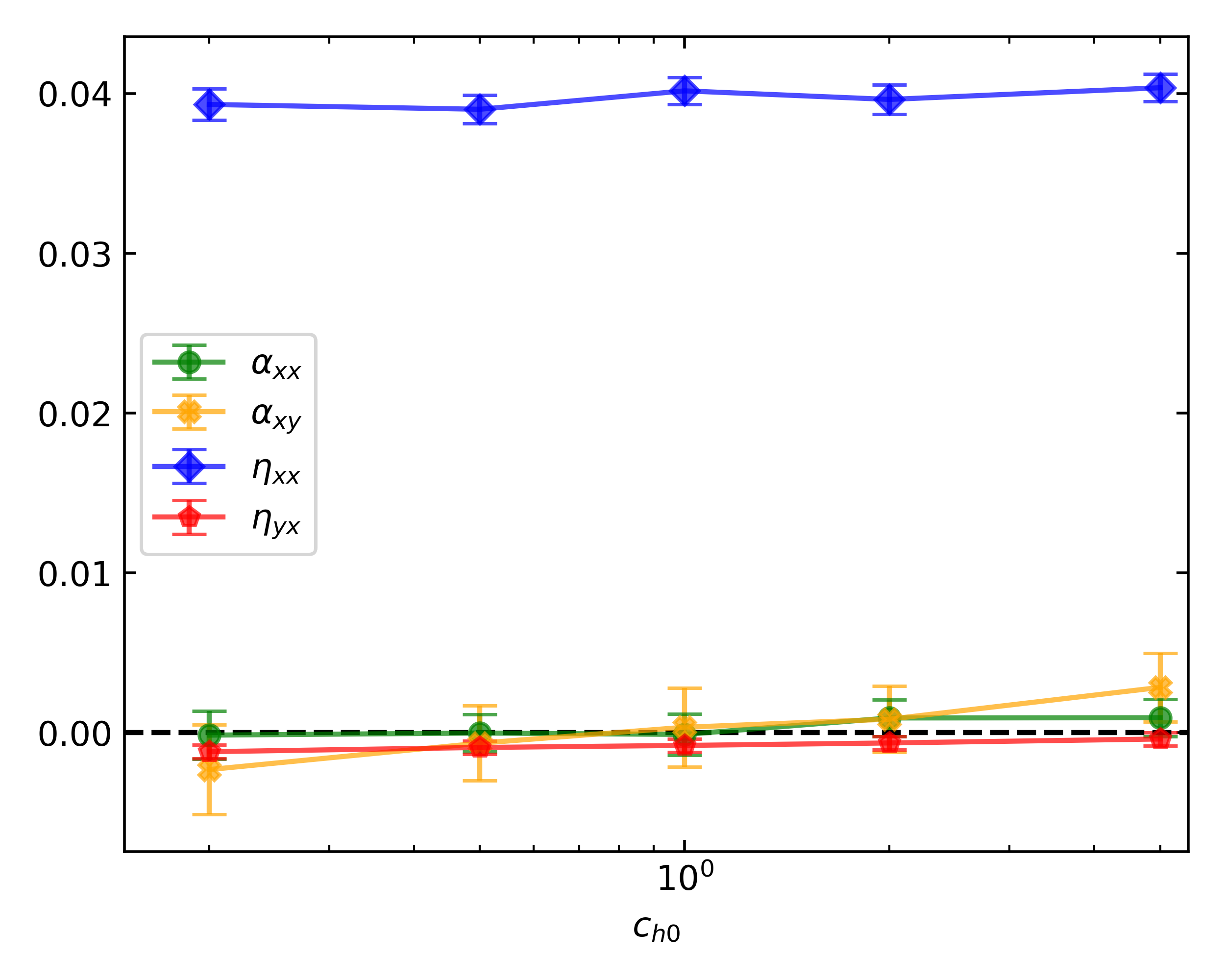}
    \caption{The time and spatially averaged transport coefficients as a function of $c_{h0}$ for unstratified simulations with background field, 48 cell per scale height resolution and box size $L_x \times L_y \times L_z = 1 \times 6.28 \times 1$. The coefficients are averaged over a period of 150 orbits starting after 50 orbits. We additionally show the statistical error of the mean value for each coefficient. As expected, $\eta_{xx}$ is the only coefficient differing significantly from zero.}
    \label{fig:transport_coeff_net_field}
\end{figure}

\begin{figure*}
    \centering
    \includegraphics[width=0.8\linewidth]{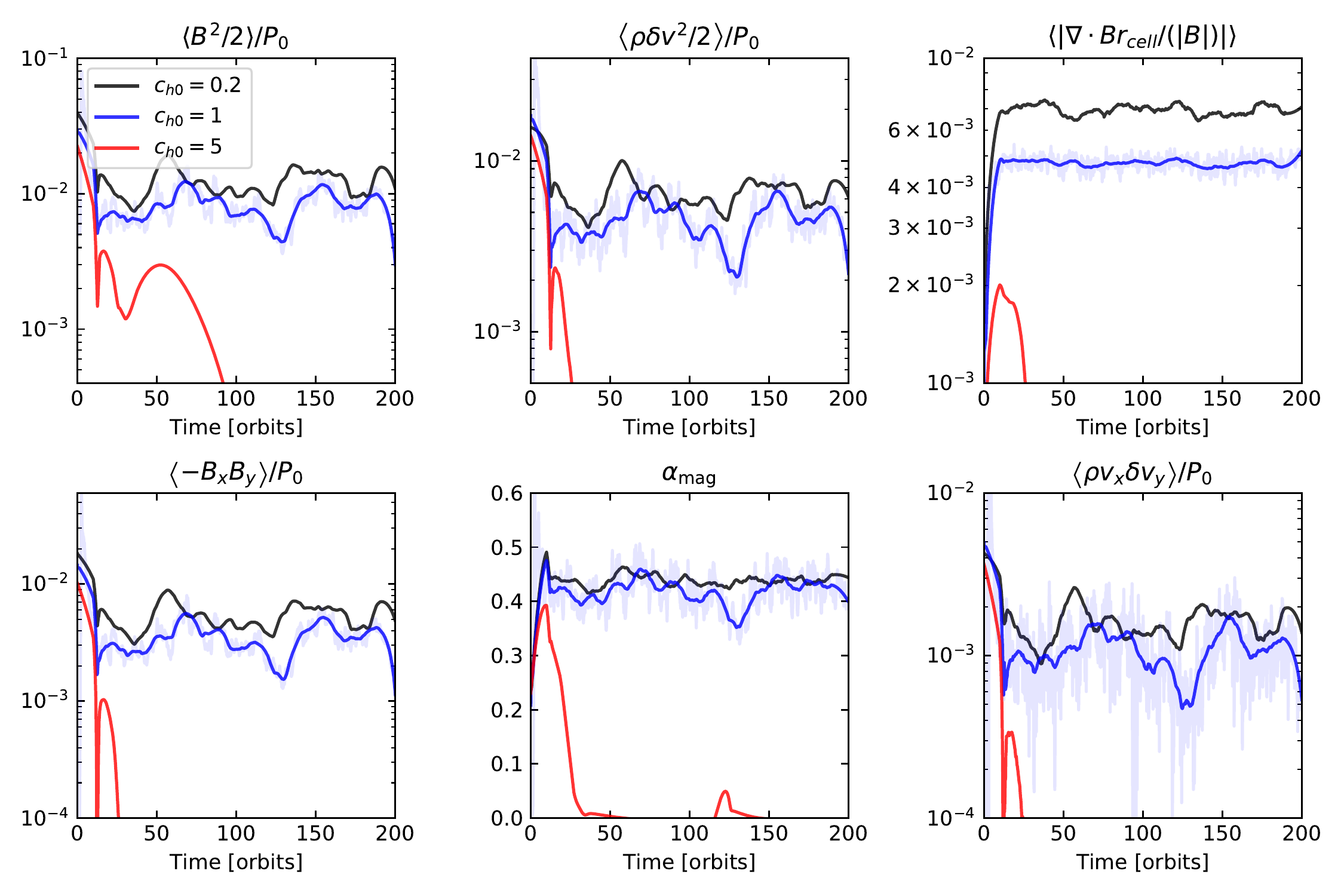}
    \caption{The temporal evolution of several volume weighted quantities for unstratified simulations without a net vertical magnetic field, box size $1\times 4 \times 1$, and $48$ cells per scale height. We vary the strength of Dedner cleaning $c_{h0}$. The shown quantities are (from left to right, and top to bottom): Magnetic field energy density, kinetic energy density, relative $\nabla \cdot \bm B$ error, Maxwell stress, normalized Maxwell stress (\ref{eq:defAlphaMag}) and  Reynolds stress.  We have smoothed the curves over 20 orbits using a Savitzky–Golay filter and show the original curve for one example case as a transparent line in the background.}
    \label{fig:evolution_small_box_zeroNetField}
\end{figure*}

\begin{figure*}
    \centering
    \includegraphics[width=0.9\linewidth]{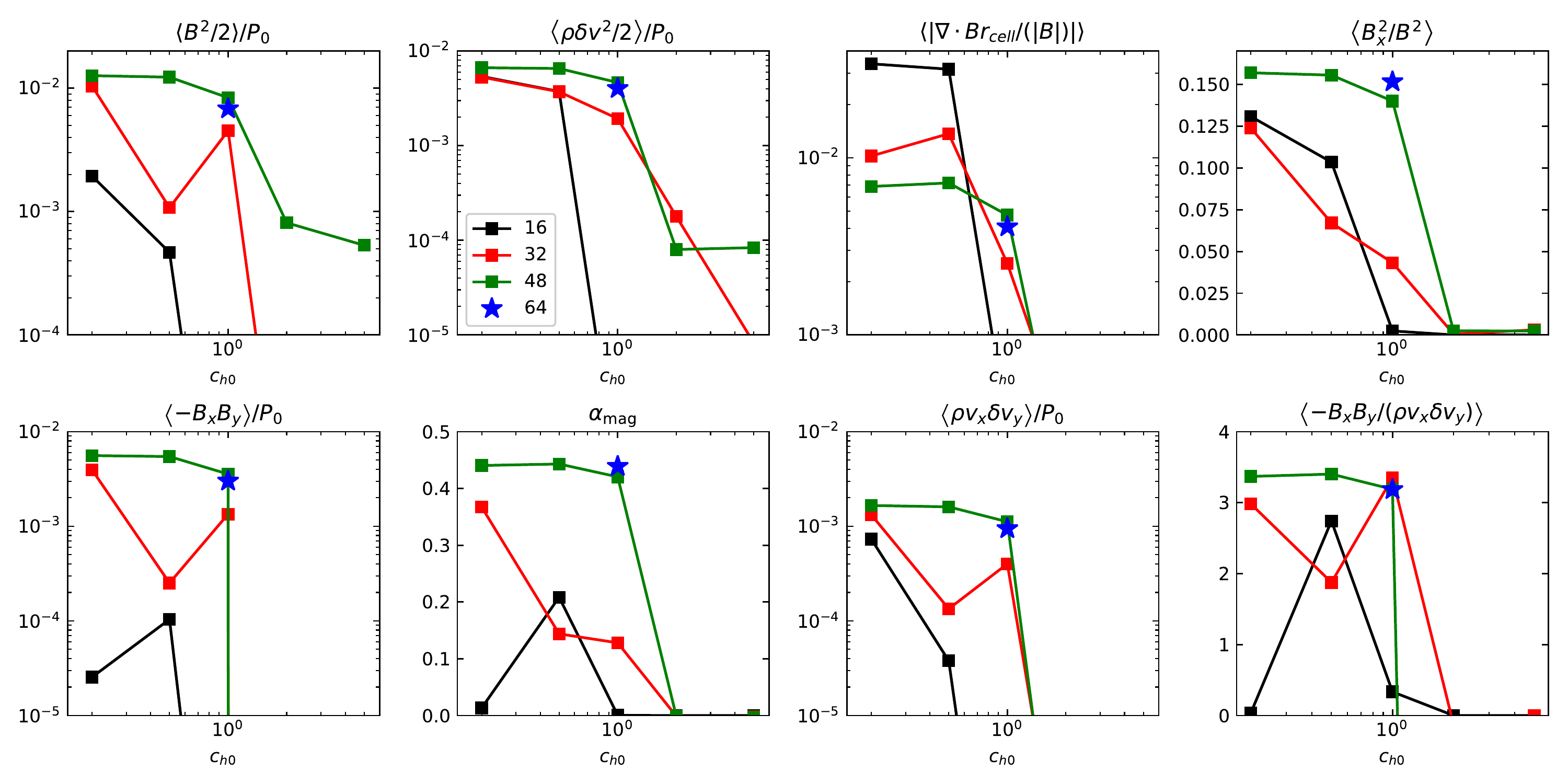}
    \caption{The temporal average of different quantities for unstratified simulations without a net vertical magnetic field and box size $1\times 4 \times 1$, as a function of the strength $c_{h0}$ of the Dedner cleaning. All quantities are averaged over 150 orbits starting after 50 orbits. We also vary the resolution with 16, 32, 48 and 64 cells per scale height, as labelled. Besides the quantities already shown in \cref{fig:evolution_small_box_zeroNetField}, we also display the ratio between radial and total magnetic field energy (top right) and the ratio of the Maxwell and Reynolds stress (bottom right).}
    \label{fig:overview_avg_zeroNetFlux_small}
\end{figure*}

\subsection{Net flux}
\label{subsec:Unstraified_net_flux}

We use a box of size $L_x \times L_y \times L_z = 1 \times 6.28  \times 1$, which corresponds to the default case of \cite{hawley1995local}, and is also discussed in \cite{deng2019local}. We set up a constant vertical magnetic field,
\begin{equation}
    \bm B = B_0 \bm \hat{e}_z,
\end{equation}
with field strength $\beta = 400$. As initial grid, we use a Cartesian mesh with $16\times 100 \times 16$ cells. For higher resolution realizations, we multiply the number of cells per dimension with a constant factor.

In \cref{fig:evolution_net_field_2pi} we show the temporal evolution of several volume-weighted properties of the MRI for simulations with 48 cells per scale height and different $c_{h0}$. In all cases, we find a saturated, turbulent state. The magnetic and kinetic energy as well as the Maxwell and Reynolds stress, and the average $\nabla \cdot \bm B$ error decrease with $c_{h0}$. This can be explained by the increasing numerical resistivity in the case of stronger Dedner cleaning.

Due to the highly time-dependent behaviour of the saturated state, we show in \cref{fig:overview_avg_NetFlux_2Pi} the volume averaged quantities additionally averaged over the last $150$ orbits (starting after 50 orbits), as a function of $c_{h0}$. We also show results for four different resolutions, although due to the computational cost we evolved the highest resolution run only for $c_{h0} = 1$. Increasing the resolution reduces the $\nabla \cdot \bm B$ error and increases the average magnetic and kinetic energy as well as the Maxwell stress. The Reynolds stress itself is more independent of the resolution and also of $c_{h0}$, while $\alpha_{\rm mag}$ only slightly increases with $c_{h0}$ and resolution. This behaviour was also observed in small box simulations by \citet{wissing2021magnetorotational}, who found for 48 cells per scale height an average value of $\alpha_{\rm mag} \approx 0.65$, which is similar to our value for $c_{h0} = 1$. Our average magnetic energy (0.2 to 0.3) as well as the Maxwell stress (0.1 to 0.2) is smaller than in \cite{hawley1995local} (0.5 and 0.3) but the Maxwell stress compares well with results from \cite{simon2008simulations} ($0.216\pm 0.116$) that used a weaker magnetic background field ($\beta = 1500$). The ratio of the Maxwell stress to the Reynolds stress is between 3 and 4, and decreases with larger resistivity.  This behaviour is similar to the one reported in \cite{wissing2021magnetorotational} and also compares favourably to \cite{hawley1995local}. In contrast, \cite{simon2008simulations} report a larger value of $7.60\pm 6.47$. Clearly, the previous results reported in the literature vary significantly, reflecting in part the turbulent behaviour of the saturated state.

As we have discussed in Section~\ref{subsec:divB}, deviations from $\nabla \cdot \bm B = 0$ can generate a net azimuthal field. We also measured this field in our simulations and did not find that it decays with resolution or smaller $\nabla \cdot \bm B $ error. But the energy associated with this net field is typically smaller by a factor of at least $10^{-4}$ compared to the average magnetic field, which is why we are confident that it does not significantly affect the general field evolution.

\cref{fig:transport_coeff_net_field} shows the average value of the transport coefficients $\alpha_{xx}$, $\alpha_{xy}$, $\eta_{xx}$ and $\eta_{yx}$. As expected, we find values close to 0 for the components of $\alpha$. Also, our measurements of $\eta_{yx}$ are compatible with 0 considering the statistical errors. Only the turbulent diffusivity $\eta_{xx}$ deviates significantly from 0 with a value of around $0.04$. These results are qualitatively similar to \cite{wissing2021magnetorotational} but our diffusivity is larger by a factor of around 5.

Smaller boxes typically lead to a stronger burst in the nonlinear regime of the MRI \citep{bodo2008aspect,lesaffre2009comparison}, since fewer active (non-axisymmetric) waves can fit in. As in \cite{deng2019local} we therefore also run some simulations in a smaller box with size $L_x \times L_y \times L_z = 1 \times 4  \times 1 $, a standard resolution of $16 \times 64 \times 16$ cells and initial $\beta = 330$. The results are very similar to the ones obtained with the the standard box, and thus we refer to Appendix~\ref{app:netFluxMRIInSmallerBox} for the corresponding figures.

\subsection{Zero net-flux}
\label{subsec:Unstraified_zero_net_flux}

A more challenging class of setups for simulation codes are the so-called zero net flux simulations. They are defined by the condition $\left< \bm B \right> = 0$, which means there is no background magnetic field that can drive the MRI. We follow the setup of \cite{deng2019local} and \cite{wissing2021magnetorotational}, and initialize a magnetic field as
\begin{equation}
    \bm B = B_0 \sin \left( 2\pi x\right) \hat{e}_z,
\end{equation}
where the initial amplitude $B_0$ is chosen such that the volume-averaged plasma $\beta$ is $\beta = 2 p_{\rm gas} /B^2 = 400$. We first run simulations in a standard box $L_x \times L_y \times L_z = 1 \times \pi \times 1$, and use initially a Cartesian grid with base resolution $16 \times 50 \times 16$ cells. We also carried out simulations with higher resolution by multiplying the number of cells per dimension with a constant factor.

In \cref{fig:evolution_small_box_zeroNetField} we show the temporal evolution of different volume-weighted quantities for a resolution of 48 cells per scale height and three different Dedner cleaning strengths. While for $c_{h0} = 5$ the MRI dies out after an initial burst we are able to sustain the MRI for at least 200 orbits for $c_{h0} \leq 1$. Additionally, we present in \cref{fig:overview_avg_zeroNetFlux_small} time-averaged values of the volume-weighted quantities as a function of $c_{h0}$ and resolution. Except for the lowest resolution calculation we find for all simulations with $c_{h0} \leq 1 $ an active MRI whereas it dies out for $c_{h0} \geq 2$. The strong dependence on the numerical resistivity (set  in our case by $c_{h0}$) of the MRI in simulations without net field and physical dissipation is also well known from the literature \citep{fromang2007mhd, deng2019local, wissing2021magnetorotational}. In particular, \cite{fromang2007mhd} showed in simulations with the finite difference code {\small ZEUS} that by increasing the resolution the MRI turbulence will be driven to smaller scales. Those scales are affected by the numerical viscosity and resistivity, and thus the final results strongly depend on numerical details.

Convergence can be regained by adding a physical viscosity and resistivity \citep{fromang2007mhdTwo}. There seems to exist a critical magnetic Prandtl number ${\rm Pr}_{m}$, which depends on the Reynolds number, below which turbulence will die out. In our case, the magnetic Prandtl number is given by the numerical viscosity and resistivity that cannot easily be measured. By increasing $c_{h0}$ the resistivity also increases, and the numerical Prandtl number decreases, which explains the existence of a critical $c_{h0}$ above which the turbulence dies out.

Experiments with SPH in \cite{wissing2021magnetorotational} found a critical Prandtl number of around ${\rm Pr}_m = 2.5$ above which the MRI turbulence survives. In contrast to static grid codes, the total stress, as well as the magnetic energy, does not decrease if we increase the resolution \citep[see e.g.][]{shi2016saturation}, which is similar to SPH for a constant magnetic Prandtl number \citep{wissing2021magnetorotational}. This might be a hint that the magnetic Prandtl number scales differently with resolution for a moving mesh code with Dedner cleaning than for a static grid code.

For our highest resolution runs, we find a total stress $\alpha \approx 0.01$ and normalized magnetic stress $\alpha_{\rm mag} \approx 0.4$, which is consistent with previous results \citep{hawley1995local, simon2008simulations, wissing2021magnetorotational}. Also, the ratio of Maxwell to Reynolds stress is in our case $\approx 3.5$ and agrees well with previous results with Eulerian codes \citep{hawley1995local, stone1996three, hawley1999local, sano2004angular}, while \cite{wissing2021magnetorotational} found values of around 4.5 with SPH. In our highest resolution run, we obtain a ratio of the radial magnetic energy to the total energy of $\left < B_x^2 / B^2 \right> \approx 0.15$, which is close to the value of 0.14 reported in \cite{shi2016saturation} and higher than $0.1$ as in \cite{wissing2021magnetorotational}.

\begin{figure*}
    \centering
    \includegraphics[width=0.8\linewidth]{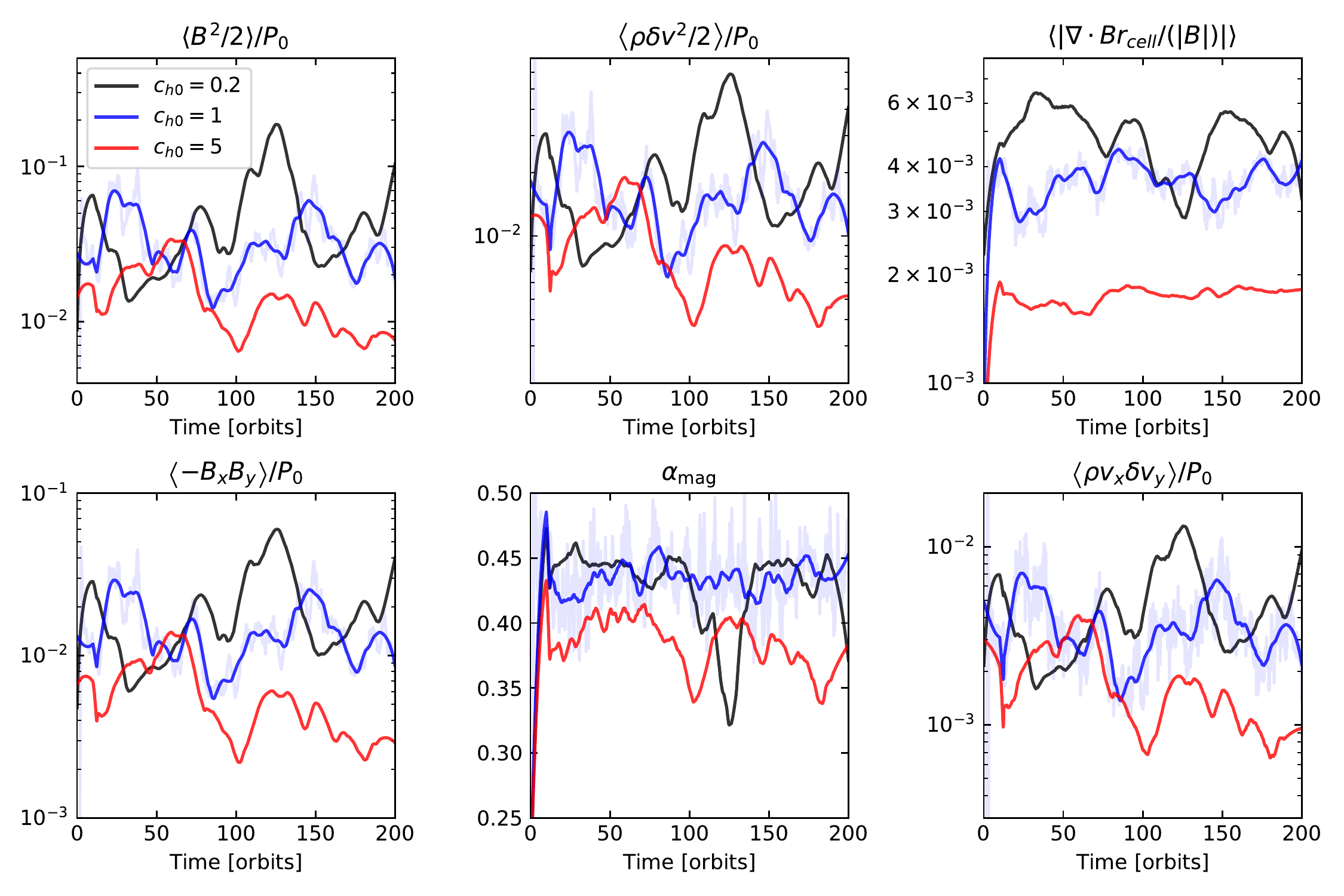}
    \caption{The temporal evolution of several volume weighted quantities for unstratified simulations without a net vertical magnetic field, with a box size $1\times 4 \times 4$, and $48$ cells per scale height. We vary the strength of Dedner cleaning $c_{h0}$, as labelled. The displayed quantities are (from left to right, and top to bottom): Magnetic field energy density, kinetic energy density, relative $\nabla \cdot \bm B$ error, Maxwell stress, normalized Maxwell stress (\ref{eq:defAlphaMag}) and Reynolds stress. We have smoothed the curves over 20 orbits using a Savitzky–Golay filter. The original measurement for one example case are included as a transparent line in the background.}
    \label{fig:evolution__tall_box_zeroNetField}
\end{figure*}

\begin{figure*}
    \centering
    \includegraphics[width=0.9\linewidth]{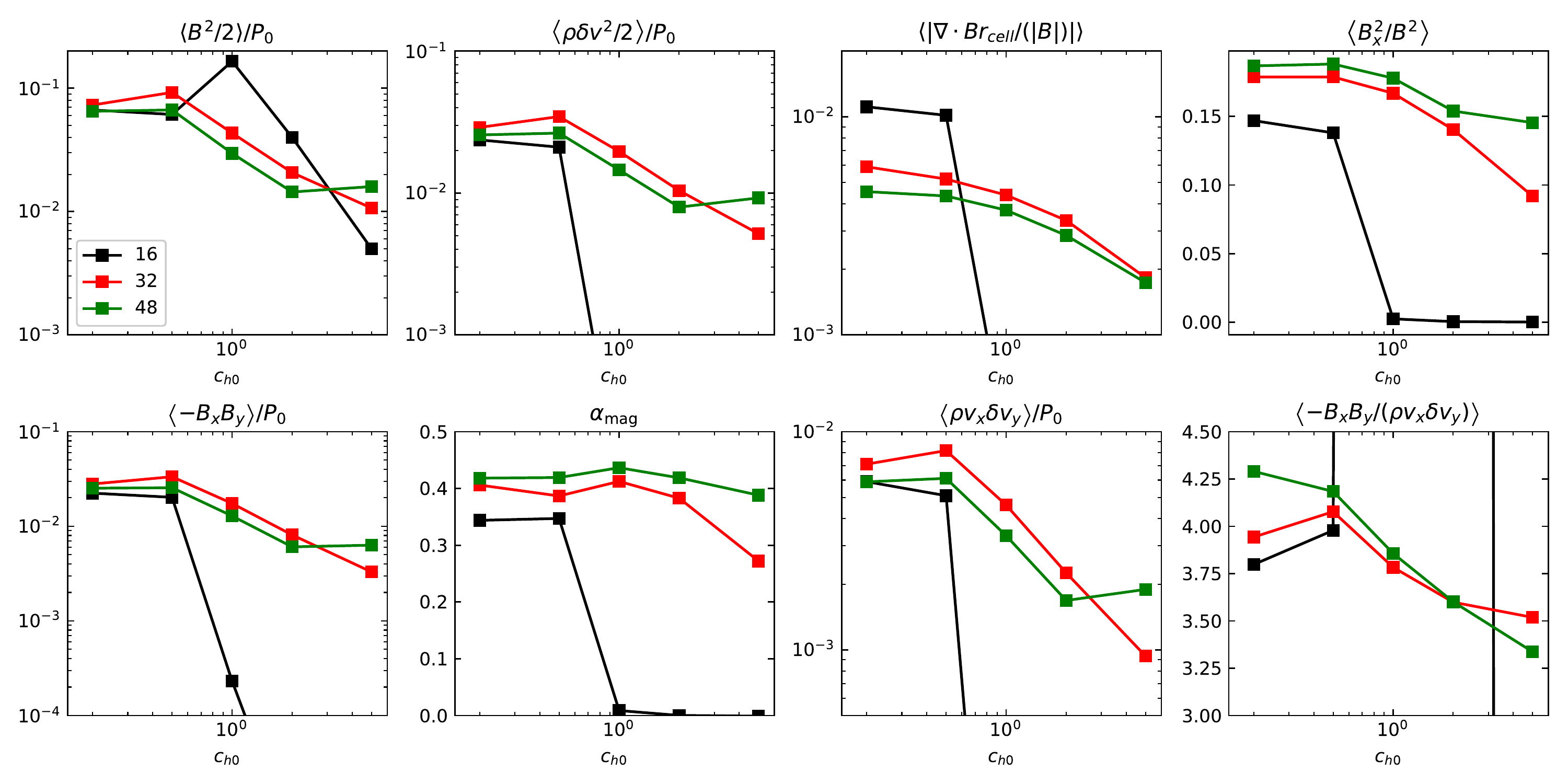}
    \caption{The temporal average of different quantities for unstratified simulations without a net vertical magnetic field, for a box size $1\times 4 \times 4$, as a function of the strength $c_{h0}$ of the Dedner cleaning. All quantities are averaged over 150 orbits starting after 50 orbits. We also vary the resolution by using 16, 32 or 48 cells per scale height, as labelled. Besides the quantities already shown in \cref{fig:evolution__tall_box_zeroNetField}, we also show the ratio between radial and total magnetic field energy (top right) and the ratio of the Maxwell and Reynolds stress (bottom right).}
    \label{fig:overview_avg_zeroNetFlux_tall}
\end{figure*}

\begin{figure*}
    \centering
    \includegraphics[width=1\linewidth]{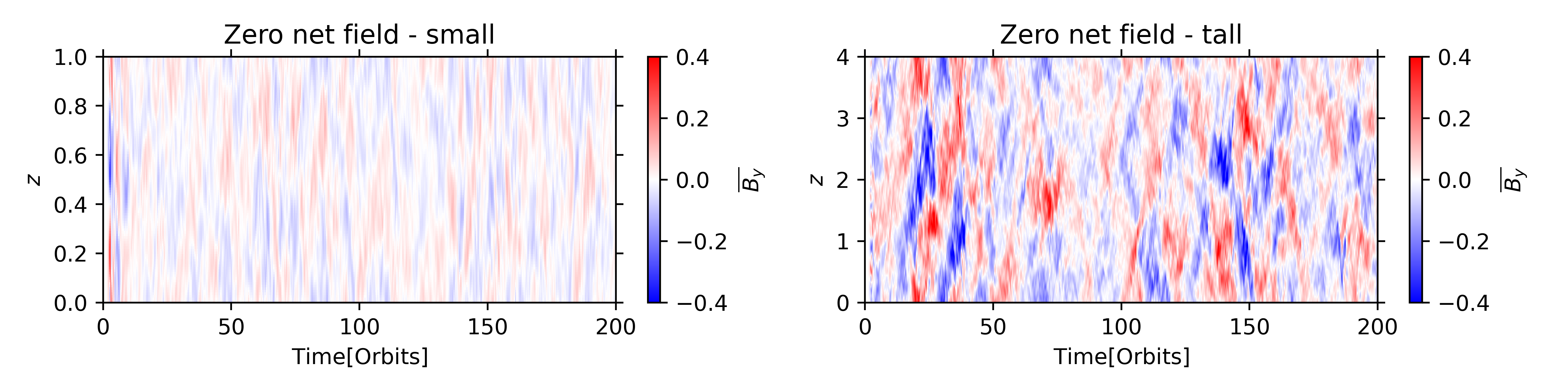}
    \caption{The evolution of the horizontally averaged azimuthal magnetic field in simulations with $c_{h0} = 1$ and 48 cells per scale height. Both simulations are evolved in an unstratified box without net field. In contrast to the small box, there is a strong large scale mean field in the tall box with comparable magnitude to the field reported in \protect\cite{shi2016saturation}.}
    \label{fig:compare_By_zeroNetField}
\end{figure*}

\begin{figure*}
    \centering
    \includegraphics[width=0.8\linewidth]{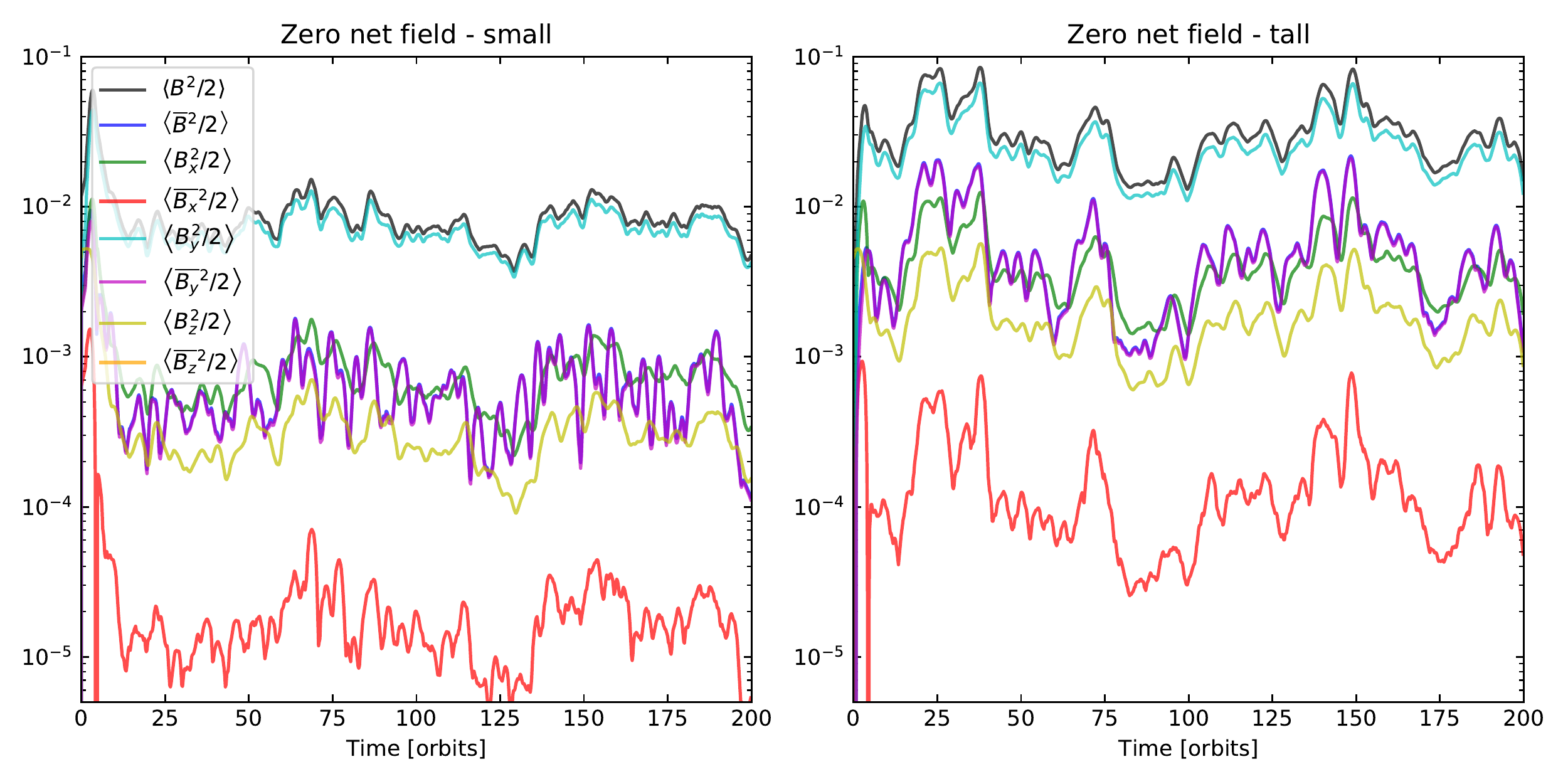}
    \caption{We show the evolution of the total magnetic energy and  contributions of different magnetic field components to it. We use simulations with  $c_{h0} = 1$ and 48 cells per scale height. Both simulations are computed in an unstratified box without net field. We note that the mean vertical field ($\bar{B}_z$) vanishes and therefore is not shown. The mean field is fully dominated by the azimuthal component.}
\label{fig:evolution_energy_zeroNetField}
\end{figure*}

\begin{figure*}
    \centering
    \includegraphics[width=0.8\linewidth]{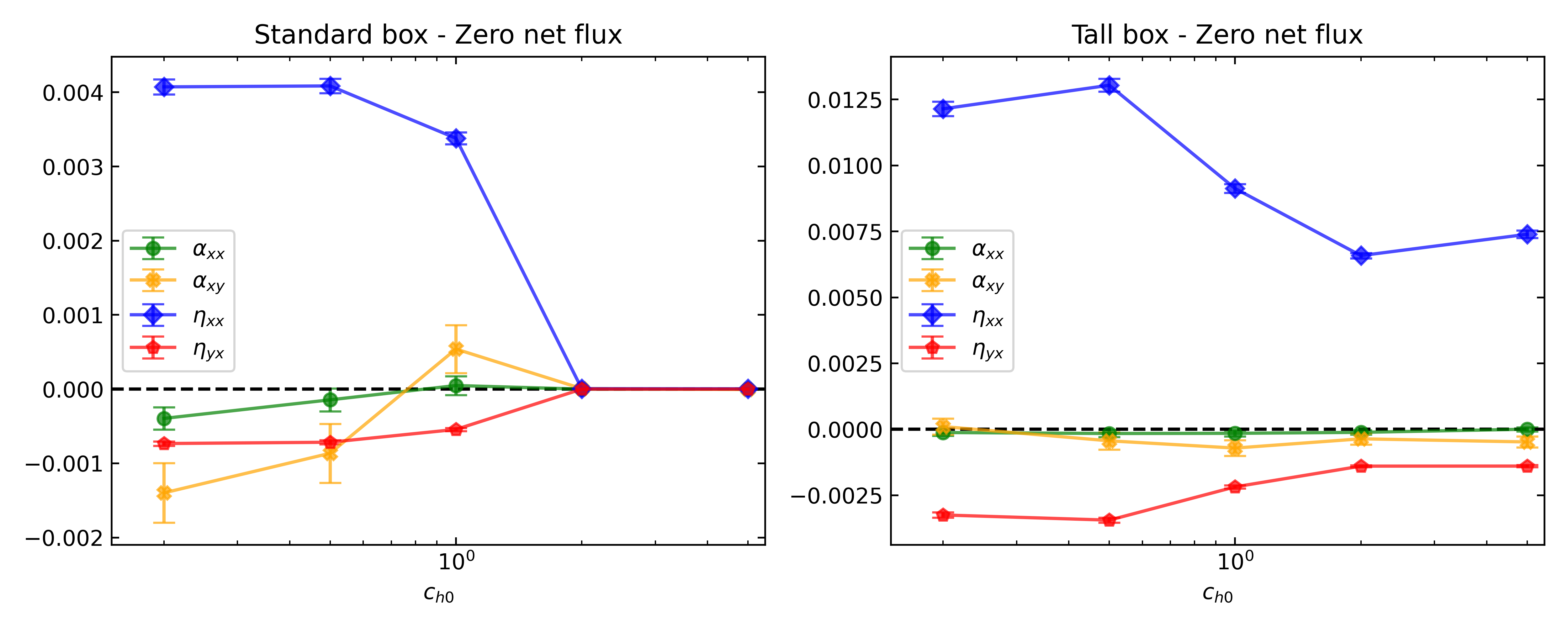}
    \caption{The transport coefficients as a function of $c_{h0}$ for unstratified simulations without background field and a resolution of 48 cells per scale height. The coefficients are averaged in space, and in time over 150 orbits starting at 50 orbits. We also show the statistical error of the mean value. As expected, $\alpha_{xx}$ as well as $\alpha_{xy}$ vanish while we find a positive turbulent diffusivity $\eta_{xx}$. The component $\eta_{yx}$ is significantly negative,  especially in the case of a tall box,  which allows the shear-current effect to be active and to generate a large scale mean field. We note that in the standard box the turbulence dies out for $c_{h0} \geq 2$.}
    \label{fig:transport_coeff_zero_net_field}
\end{figure*}

\subsection{Zero net-flux simulations in tall boxes}
\label{subsec:Unstraified_zero_net_flux_tall}

In boxes with larger vertical aspect radio ($L_z / L_x \geq 2.5$) a new and more vigorous MRI dynamo emerges. \cite{shi2016saturation} showed that in this case the stress becomes independent of the resolution, which simplifies the comparison of results of different codes. We therefore rerun the simulations from the previous subsection in a larger box, $L_x \times L_y \times L_z = 1 \times 4 \times 4$, with a base resolution of $16\times 64 \times 64$ cells, and using the same initial field as in the smaller box.

In \cref{fig:evolution__tall_box_zeroNetField} we show the temporal evolution of volume-weighted quantities for a resolution of 48 cells per scale height and different Dedner cleaning speeds. In contrast to the smaller box, the MRI can sustain turbulence even for $c_{h0} =5$. As one can see in \cref{fig:overview_avg_zeroNetFlux_tall}, only for the lowest resolution and $c_{h0} \geq 1$ the MRI dies out, while in general the stress is larger by a factor of more than 4 compared to the small box. The magnetic energy and also the Maxwell and Reynolds stress have in our higher-resolution simulation a maximum at $c_{h0} = 0.5$ and decrease with stronger numerical resistivity. The dependence of the saturated quantities on $c_{h0}$ is in this case stronger than for simulations with a net field.

The results for $c_{h0} =0.5$ and the ones from \cite{shi2016saturation} compare in general very well for the same resolution of 32 cells per scale height.  We find a total stress of $\alpha \approx 0.042$, whereas \cite{shi2016saturation} measure $\alpha \approx 0.052$. But it seems that our results are already converged with higher resolution while in their paper the stress further increases with higher resolution. In the case of the tall box, we find as expected a strong azimuthal mean field whose evolution is shown in \cref{fig:compare_By_zeroNetField}. The amplitude is higher than in \citet[][see their Fig. 4]{wissing2021magnetorotational} and compares  qualitatively well with the results in \citet[][see their Fig. 10]{shi2016saturation}. 

In \cref{fig:evolution_energy_zeroNetField} we also show the distribution of the magnetic energy over the different spatial components and subdivide it by the mean and fluctuating parts. The total magnetic energy is dominated by the azimuthal component with a contribution of around 10\% from the radial component. The energy of the mean magnetic field is fully dominated by the azimuthal component, which is as expected larger in the tall box simulation in comparison to the small box simulation. The mean azimuthal field is still smaller in comparison to the simulations in \cite{shi2016saturation}, where it contributes around 50\% of the magnetic energy. Times of lower magnetic energy in the mean-field component are also visible in the space diagrams in \cref{fig:compare_By_zeroNetField}, where a large-scale magnetic field is missing (e.g.~at around 80 orbits).

In \cref{fig:transport_coeff_zero_net_field} we show the spatially and temporally averaged transport coefficients. As expected, all components of $\alpha$ are close to 0 while we find both in the small and the tall box a significantly positive value for the turbulent diffusivity $\eta_{xx}$. Its value is around $0.003$ for the standard box in the cases with sustained turbulence, and somewhat larger in the tall box. We also find that $\eta_{yx}$ is slightly negative in the standard box, while  it becomes more significantly negative in the tall box. These findings are consistent with the results of \citet{shi2016saturation} but opposite to those of \citet{wissing2021magnetorotational} who found slightly positive values for the tall box case and vanishing values for the standard box case. This suggests that the shear-current effect can be followed in our simulations, explaining why we get a significant large-scale magnetic field as in \cite{shi2016saturation}, in contrast to \cite{wissing2021magnetorotational}. Our result for the magnitude of $\eta_{yx}$ compare well with the results from \cite{shi2016saturation}, who equally found $\eta_{yx} \approx -10^{-3}$.

\section{Stratified simulations}
\label{sec:stratifiedSimulation}

In this section, we present simulations that include the linearized vertical component of the gravitational force of the central object as presented in equation~(\ref{eq:shearingBoxSourceTerms}). For an isothermal gas the hydrostatic density profile is given by
\begin{equation}
    \rho(z) = \rho_0 \exp \left(- \frac{z^2}{2H^2}\right) ,
    \label{eq:densStratified}
\end{equation}
with the scale height\footnote{We note that some studies add a factor $\sqrt{2}$ in the definition of the scale height \citep[e.g.][]{simon2011resistivity}.} $H= c_s/ \Omega_0$  and the mid plane density $\rho_0 = 1$. To allow for outflows we require a relatively large $L_z$. But in this case $\rho(z)$ becomes very small which
can lead to numerical problems, which is why we introduce a density floor $\rho_{\rm min} = 10^{-6}$ and replace the acceleration in the vertical direction by
\begin{equation}
    a= \frac{\rho - \rho_{\rm min}}{\rho} \Omega_0^2\, z.
\end{equation}
After each time step, we set the density to $\rho_{\rm min}$ for cells with $\rho < \rho_{\rm min}$ and keep the velocity and the magnetic field fixed.

{\small AREPO} allows the creation and destruction of new cells (refinement/derefinement in the following) during  run time. These processes can be triggered by more or less arbitrary criteria, which are typically based on the mass and volume of the cells. In this section we define a target mass $m_{\rm target}$ and split cells with a mass higher than $2\, m_{\rm target}$, and remove cells with a smaller mass than $0.5\, m_{\rm target}$. To avoid too rapid local variations in the spatial resolution we impose a maximum allowed volume ratio of 10 between adjacent cells, and enforce a maximum volume of $0.05$ per cell. The volume-based conditions become especially important in the low-density halo of the disk and avoid that a cell interacts with a periodic image of itself.

Without magnetic field the density profile (\ref{eq:densStratified}) in combination with the background velocity shear profile from equation~(\ref{eq:backroundShearFlow}) should be stable. As in \cite{deng2019local} and \cite{wissing2021magnetorotational} we use a box of size $L_x \times L_y \times L_z = \sqrt{2} \times 4 \sqrt{2} \times 24$
and an initial magnetic field
\begin{equation}
    \bm B = B_0 \hat{e}_y,
\end{equation}
with plasma $\beta = 25$. We have run simulations with $c_{h0} = 0.5$ and  $c_{h0} =1$, combined with $m_{\rm target} = 10^{-5}$ (approx. $1.6\times 10^6$ cells) or $m_{\rm target} = 5 \times 10^{-6}$ (approx. $3.2\times 10^6$ cells).

\begin{figure*}
    \centering
    \includegraphics[width=1\linewidth]{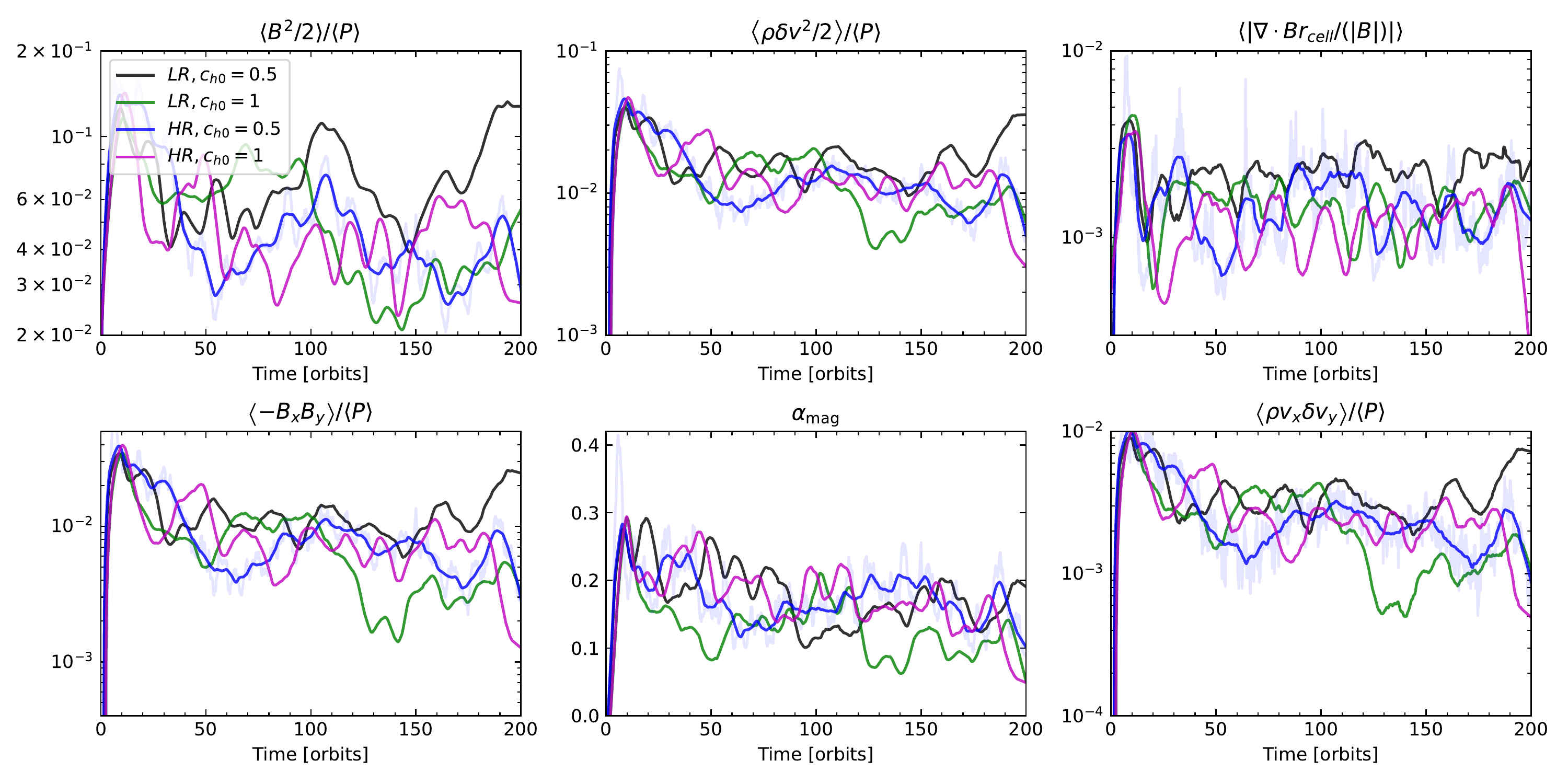}
    \caption{The temporal evolution of several volume weighted quantities for stratified simulations with box size $\sqrt{2}\times \sqrt{2} \times 24$, for two different resolutions (LR:  $m_{\rm target} = 10^{-5}$, HR: $m_{\rm target} = 5\times 10^{-6}$) and two different strengths of the Dedner cleaning $c_{h0}$. The shown quantities are (from left to right, and top to bottom): Magnetic field energy density, kinetic energy density, relative $\nabla \cdot \bm B$ error, Maxwell stress, normalized Maxwell stress (\ref{eq:defAlphaMag}) and Reynolds stress. The quantities are averaged over the whole simulation box. We have smoothed the curves over 10 orbits using a Savitzky–Golay filter, and show the original curve for one example case as a transparent line.}
    \label{fig:evolution_stratified}
\end{figure*}

\begin{figure*}
    \centering
    \includegraphics[width=1\linewidth]{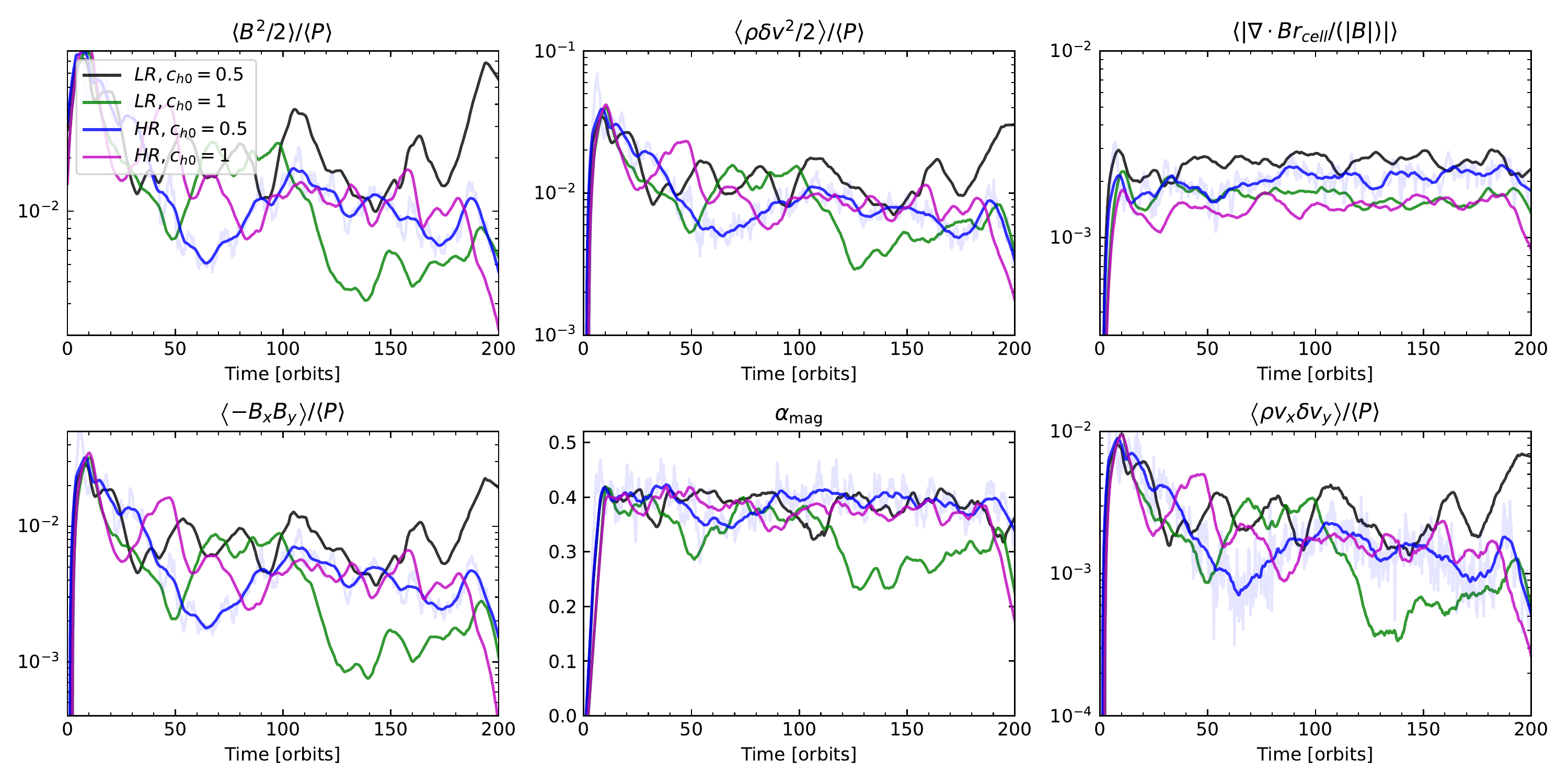}
    \caption{Same as \cref{fig:evolution_stratified}, but we only consider gas close to the mid plane ($\pm \sqrt{2} H$).}
    \label{fig:evolution_stratified_high_dens}
\end{figure*}

\begin{figure*}
    \centering
    \includegraphics[width=1\linewidth]{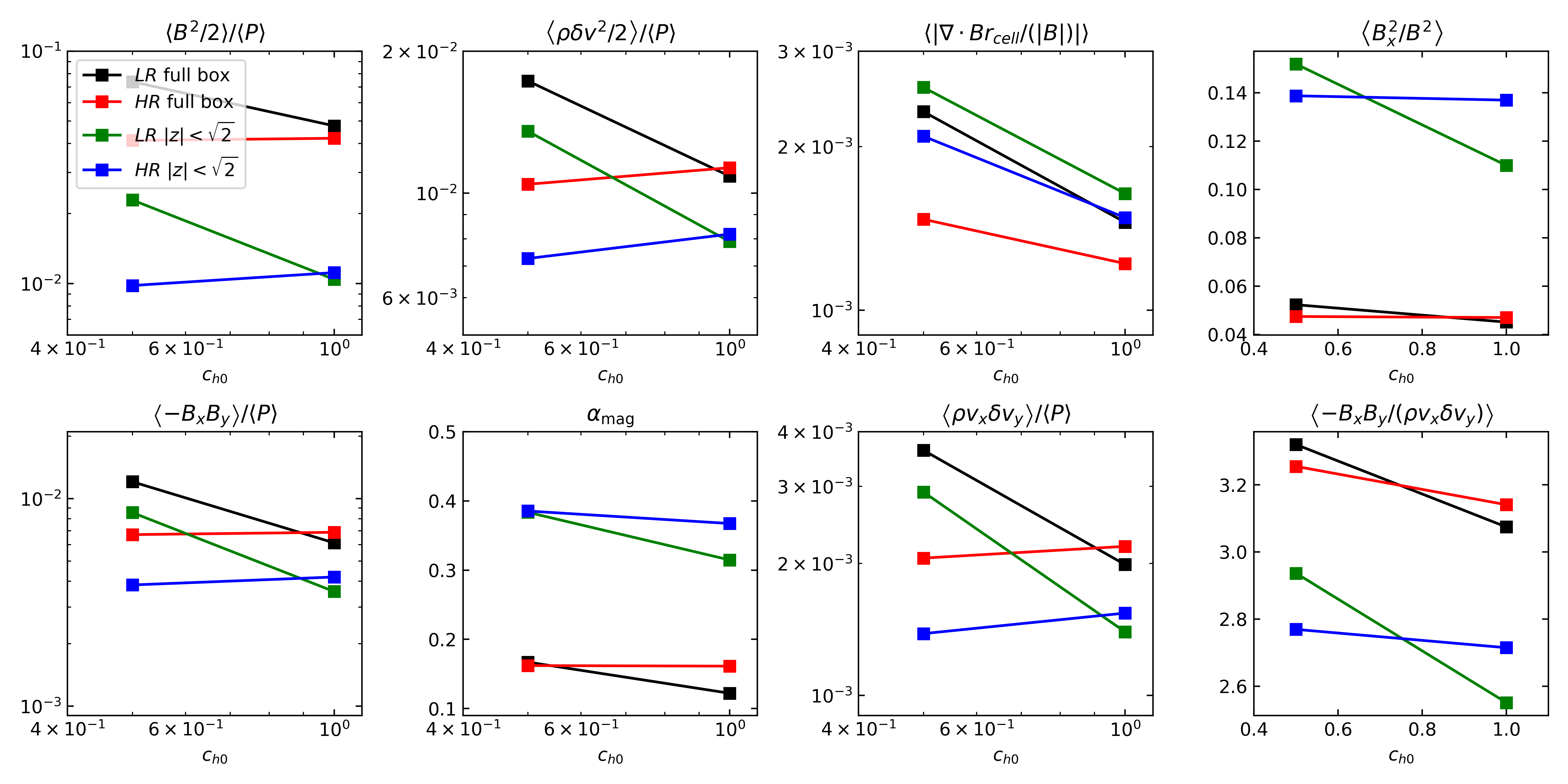}
    \caption{The temporal average of different quantities for stratified simulations with box size $\sqrt{2}\times \sqrt{2} \times 24$ as a function of the strength $c_{h0}$ of the Dedner cleaning. All quantities are averaged over 150 orbits starting after 50 orbits.  We also vary the resolution (LR:  $m_{\rm target} = 10^{-5}$, HR: $m_{\rm target} = 5\times 10^{-6}$). Besides the quantities already shown in \cref{fig:evolution_stratified} and \cref{fig:evolution_stratified_high_dens} we also include the ratio between radial and total magnetic field energy (top right), and the ratio of the Maxwell and Reynolds stress (bottom right).}
    \label{fig:temporal_avg_stratified}
\end{figure*}

In \cref{fig:evolution_stratified} we show the temporal evolution of several volume averaged quantities for our four runs averaged over the whole box, while in \cref{fig:evolution_stratified_high_dens} we show the corresponding plot only for cells close to the mid plane ($\pm \sqrt{2} H)$. In this volume the MRI is active, so that this filter avoids averaging over a magnetically dominated corona. In all simulations, the magnetic field gets amplified at the onset of the MRI and saturates later on into a turbulent state. 
Although the absolute energy, as well as the stresses, are dominated by the high-density region, they are larger in the low-density region after normalizing them with the average pressure.

In \cref{fig:temporal_avg_stratified} we show additionally the temporal average of different quantities as a function of $c_{h0}$. As expected, the $\nabla \cdot \bm B$ error decreases with $c_{h0}$ and also with higher resolution. The magnetic field and stress decrease for larger $c_{h0}$ for the low-resolution simulation, while for the high-resolution simulation the results seem to be approximately independent of $c_{h0}$. The total stress is around $\alpha \approx 0.005 -0.012$, and the normalized stress is $\alpha_{\rm mag} \approx 0.3 - 0.4$ close to the mid plane region, which agrees well with the results in \cite{wissing2021magnetorotational} for GDSPH  with an artificial viscosity parameter $\alpha_B = 0.5$. Also, the magnetic energy density is similar, but we find a ratio of Maxwell to Reynolds stress of $\approx 2.6 -3$, in contrast to the value $\simeq 4$ found by \cite{wissing2021magnetorotational}.
\cite{hawley2011assessing} compared several previous stratified MRI simulations run with static grid codes \citep{simon2011resistivity,shi2009numerically,davis2010sustained,guan2011radially} and found for the normalized stress $\alpha_{\rm mag} \approx 0.22 - 0.4$, for the total stress $\alpha \approx 0.01 - 0.03$ and the ratio of the square of the radial to total magnetic field $\left<B_x^2 / B^2 \right> \approx 0.07- 0.18$ close to the mid plane ($\pm \sqrt{2} H$ for our definition of the scale height).
\citet{deng2019local} found during the time period with active MRI similarly $\alpha_{\rm mag} \approx 0.4$ and $\alpha \approx 0.02 - 0.04$, which were  calculated with a mass-weighted average.
While our values for $\alpha_{\rm mag}$ compare well with those previous studies our total stresses are at the lower end of the results previously reported in the literature.

\begin{figure*}
    \centering
    \includegraphics[width=1\linewidth]{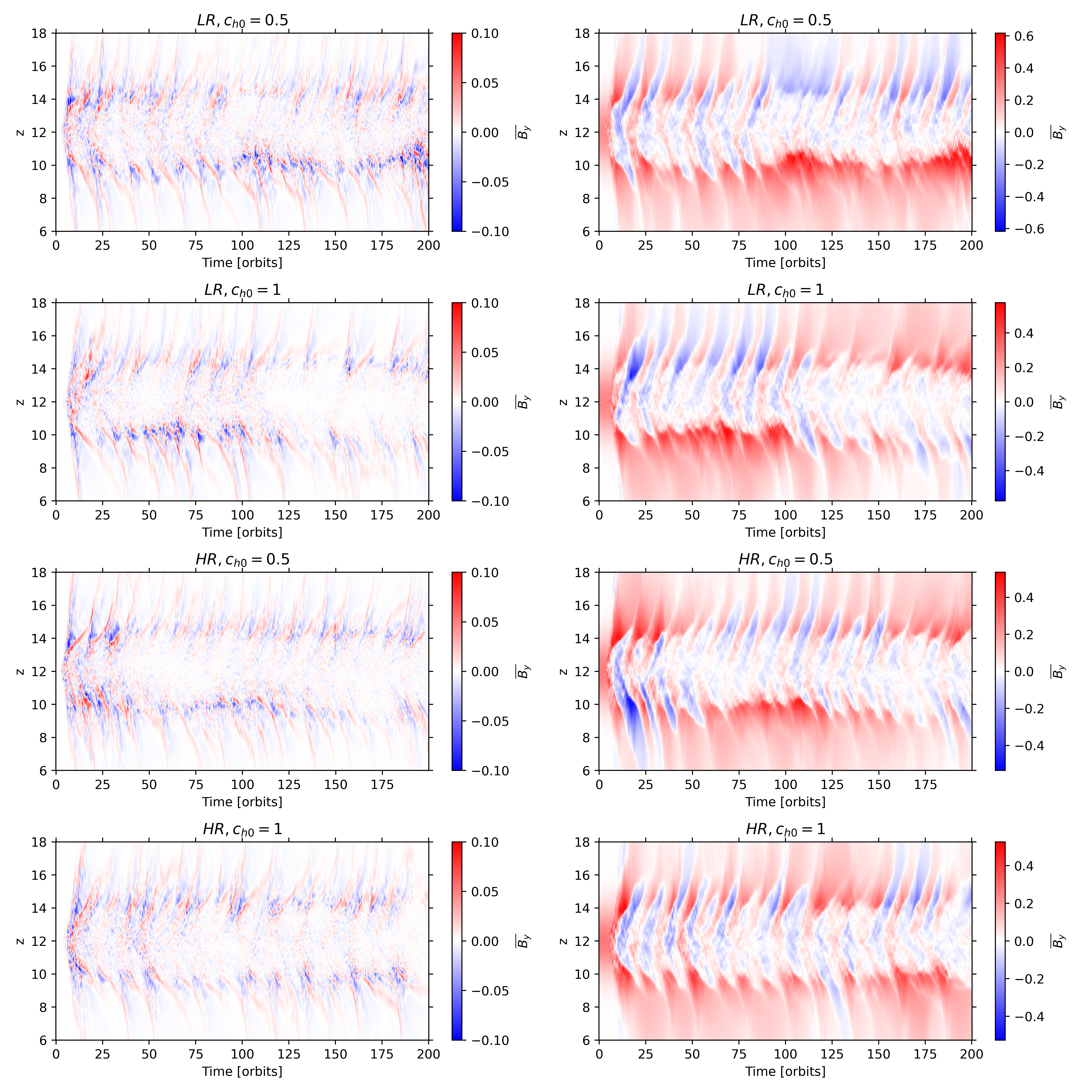}
    \caption{Space-time diagram of the horizontally averaged radial (left) and azimuthal (right) magnetic field for our four stratified simulations. In all simulations, we see remnants of the characteristic butterfly diagram, although the outer regions are dominated by a positive azimuthal net field.}
    \label{fig:horizontalAverage_stratified}
\end{figure*}

\begin{figure*}
    \centering
    \includegraphics[width=0.9\linewidth]{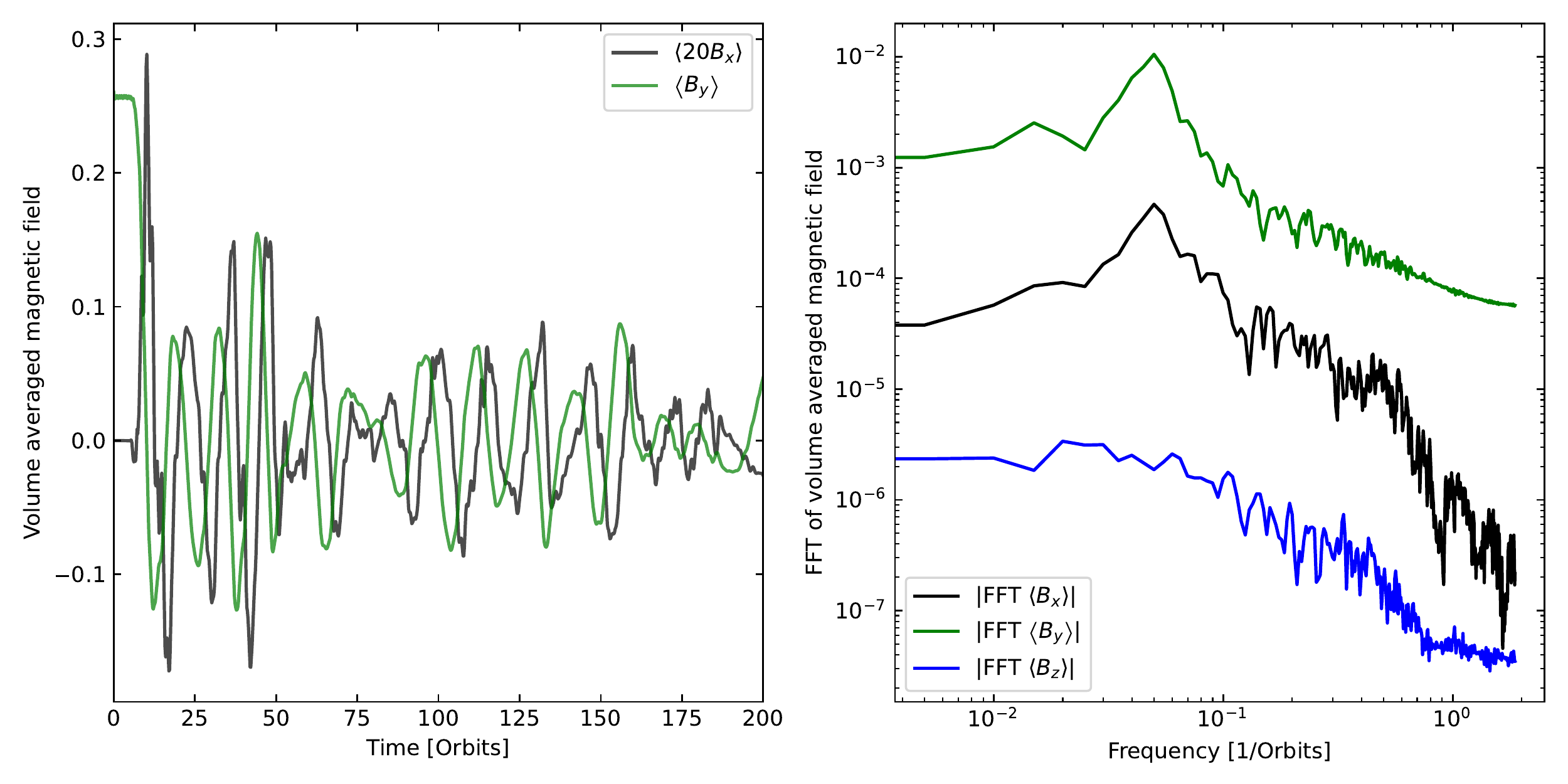}
    \caption{{\em Left panel:} The temporal evolution of the volume-averaged radial and azimuthal magnetic fields near the mid-plane ($\pm H$) for the HR simulation with $c_{h0} = 1$. For visual clarity, we multiplied $B_x$ with a factor of 20. As also seen in \cref{fig:horizontalAverage_stratified}, the mean field is oscillating with a period of around 14 orbits. There is a time lag between the radial and azimuthal fields which can be explained with an active $\alpha \omega$ dynamo and which is consistent with results from \protect\cite{simon2011resistivity}. {\em Right panel:} Temporal power spectrum of the signal on the left panel, started after 50 orbits. One can see a peak for the radial and azimuthal fields for the oscillation frequency of the butterfly diagram.}
    \label{fig:mean_field_evolution}
\end{figure*}

As one can see in \cref{fig:horizontalAverage_stratified}, all our simulations show the characteristic butterfly diagram in the horizontally averaged magnetic field. The azimuthal magnetic field gets buoyantly transported out of the mid-plane and flips sign in the central region. This behaviour can be observed over 100s of orbits in the inner region, whereas in the outer regions a positive azimuthal field forms in most simulations. In \cref{fig:mean_field_evolution} we show the volume-averaged mean azimuthal field close to the mid-plane and its temporal Fourier transformation. We find in all simulations an average period of 14, which is consistent with the results from \cite{simon2011resistivity}.

\begin{figure*}
    \centering
    \includegraphics[width=0.9\linewidth]{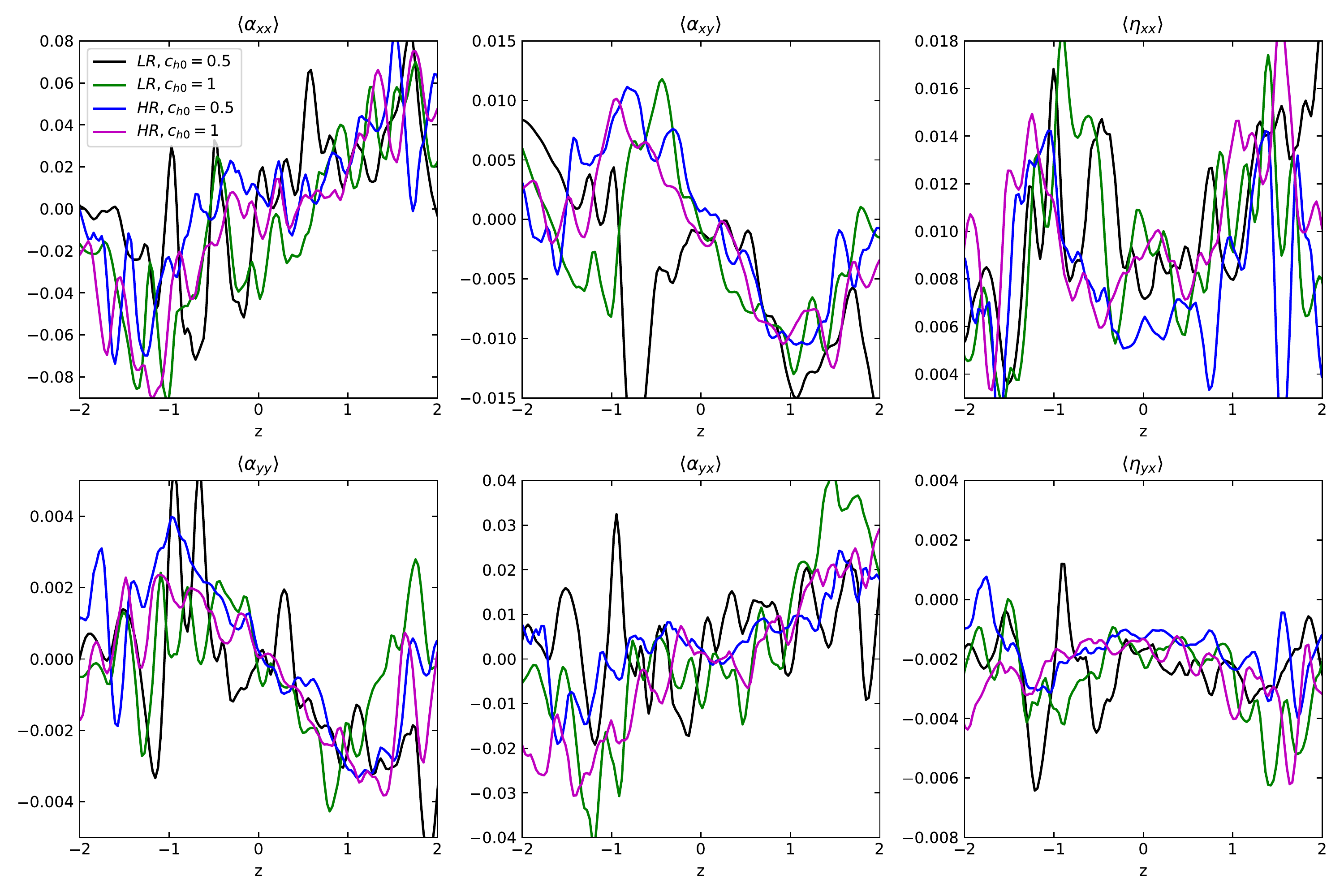}
    \caption{Temporally averaged transport coefficients as a function of $z$ for stratified simulations with a net field. We averaged over 150 orbits starting after 50 orbits. Except for the low resolution simulation with $c_{h0} = 0.5$, all simulations show clear gradients in the four components of $\alpha$ close to the mid-plane at $z=0$.  We find an active $\alpha \omega$ dynamo (due to negative gradients of $\alpha_{yy}$) and a transport of the mean field away from the mid-plane (due to a positive gradient in the diamagnetic pumping term $\gamma_z$, see equation~\ref{eq:diamagneticPumpingTerm}).  The turbulent diffusivity $\eta_{xx}$ is positive, and the shear-current effect is active due to $\eta_{yx} <0$.}
    \label{fig:transport_coeff_z_distribution_stratified}
\end{figure*}

There is a time lag between the radial and azimuthal fields which is characteristic of an active $\alpha\omega$ dynamo. A positive net radial field will decrease the net azimuthal field following the first term in equation~(\ref{eq:dBdty}). The net azimuthal field becomes negative and starts to damp the radial field following the second term in equation~(\ref{eq:dBdtx}) if $\alpha_{yy}$ has a negative gradient in the $z$-direction. In \cref{fig:transport_coeff_z_distribution_stratified} we, therefore, show the horizontally and temporally averaged transport coefficients as a function of $z$. Except for the low resolution run with $c_{h0} =0.5$, we find for all four components of $\alpha$ clear gradients close to the mid-plane. All of them are antisymmetric with respect to the mid-plane position, as expected. $\alpha_{yy}$ has a negative gradient and can therefore create and amplify the radial magnetic field if there is a mean azimuthal field as given in our simulation. This means the $\alpha \omega$ dynamo is active in our simulations, a result that is consistent with previous findings in the literature \citep{brandenburg1995dynamo, brandenburg2002local, brandenburg2008dual, shi2009numerically, wissing2021magnetorotational}. $\alpha_{xx}$ has a positive gradient and therefore counteracts the rotational term for the evolution of $\overline{B_y}$, but the latter one still dominates. Finally, $\alpha_{xy}$ has a positive gradient while $\alpha_{yx}$ has a negative gradients, which means the diamagnetic pumping term $\gamma_z = 0.5 \left(\alpha_{yx}-\alpha_{xy}\right)$ is positive above the mid-plane and negative below. The mean magnetic field, therefore, gets transported away from the mid-plane in our simulations, which is consistent with the results in \cite{shi2016saturation} and \cite{wissing2021magnetorotational}. Also, the absolute amplitude of $0.01$ to $0.02$ compares well with \cite{shi2016saturation} in contrast to \cite{wissing2021magnetorotational}, who found $|\alpha_{yx}| \ll |\alpha_{xy}|$.

We find a positive turbulent diffusivity $\eta_{xx} \approx 0.01$ consistent with the results from \cite{shi2016saturation} and \cite{wissing2021magnetorotational}, with $\eta_{yx} < 0$ close to the mid plane. The latter result shows that  the shear-current effect is active in our simulations, as in \cite{shi2016saturation}, but unlike in the results of \cite{wissing2021magnetorotational}. This is similar to our findings for the unstratified, tall box simulation without a mean magnetic field. The quantitative value of $\approx -2\times 10^{-3}$ also agrees well with \cite{shi2016saturation}.

\begin{figure*}
    \centering
    \includegraphics[width=0.9\linewidth]{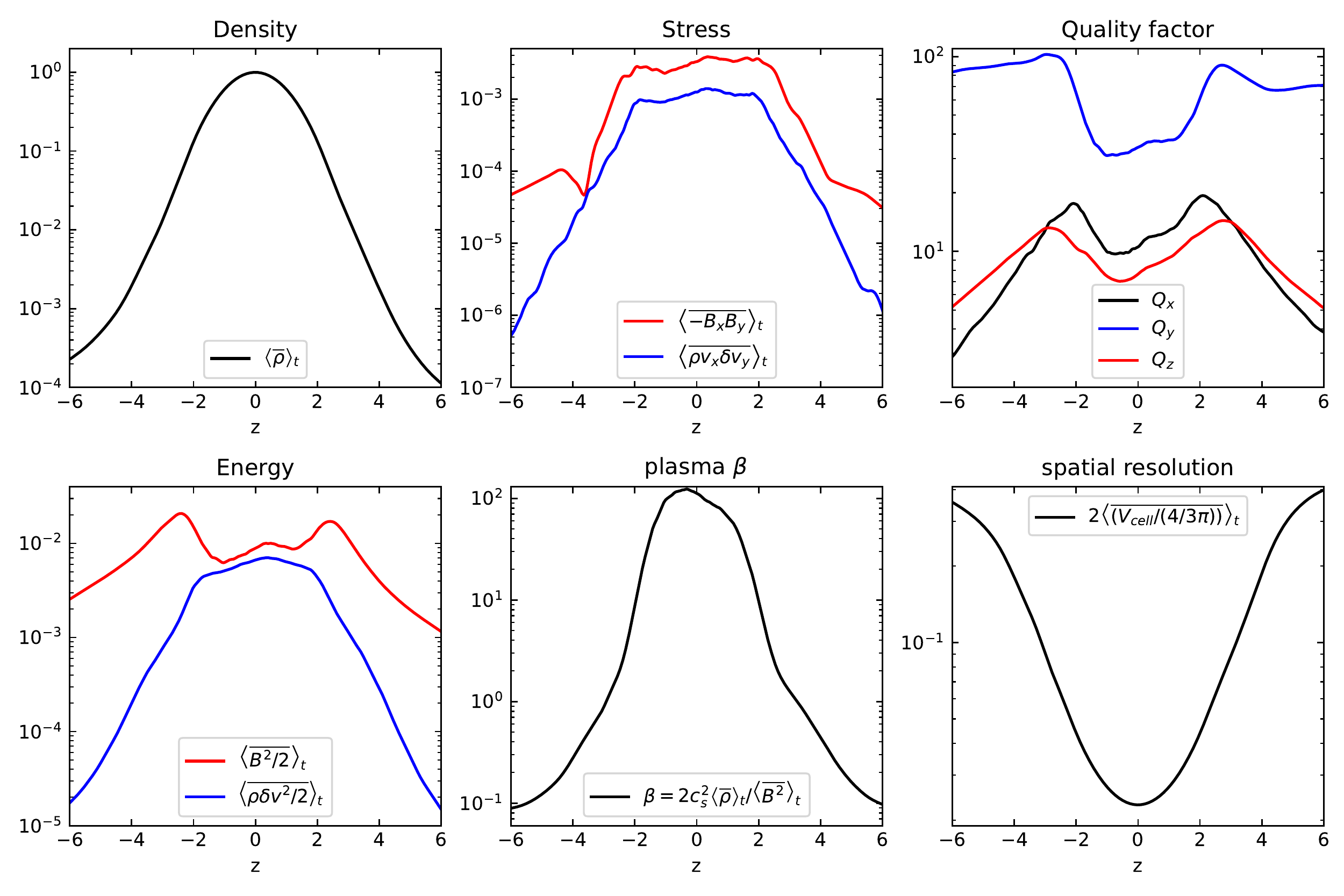}
    \caption{Horizontally and temporally averaged vertical profiles of different properties for the HR stratified simulation with $c_{h0} = 1$.
    From the left to right in the upper panel, we show the profiles of the density, stress, quality factor, energy, plasma $\beta$, and average cell diameter.}
    \label{fig:z_profile_averages}
\end{figure*}

In Figure~\ref{fig:z_profile_averages}, we show additionally several vertical profiles for the HR simulation with $c_{h0} = 1$.  In the region within a distance $\pm 2H$ from the mid-plane, we find an approximately constant Maxwell and Reynolds stress as well as uniform energy density. At the boundary of this region, $\beta$ reaches unity and the system becomes magnetically dominated further away from the mid-plane. In the outer region, the turbulent, kinetic energy drops faster than the magnetic energy and the system is stable to the MRI.

This all agrees qualitatively well with the results in \cite{simon2011resistivity} though we find a bump in the magnetic energy at the boundary between the MRI and magnetic field-dominated domains. This can also be observed in the butterfly diagrams in \cref{fig:horizontalAverage_stratified}. Close to the mid-plane we always find $Q_z > 7$, $Q_x > 10$, and $Q_y > 30$, which is close to the condition $Q_z > 10$ and $Q_y > 20$ to reach convergence in the stresses \citep{hawley2011assessing}. The spatial resolution is, as expected, highest in the high-density region close to the mid-plane.

In summary, our simulations compare qualitatively well with previous simulations, though the saturated stresses and magnetic energy seem to lie at the lower end of reported results in the literature \citep{hawley2011assessing, deng2019local,wissing2021magnetorotational}. This could be explained by the problem that in our simulations the Dedner cleaning speed is given by the maximum signal speed in the system, which we typically find in the corona of our disk.  The maximum signal speed is on average 10 times larger than the sound speed (see \cref{fig:signalSpeedEvolution}), which is close to the signal speed in the mid-plane. This means that we typically clean the magnetic field much more strongly in the mid-plane as would be required locally, and therefore the numerical resistivity is also (needlessly) larger. Smaller values for $c_{h0}$ would solve this problem but in this case the magnetic field evolution in the corona tends to become unstable.

\section{Summary and Conclusions}
\label{sec:discussionSummary}

In this paper, we analyzed the ability of the moving-mesh code {\small AREPO} to simulate the linear and nonlinear stages of the magnetorotational instability using the shearing box approximation. The code can accurately resolve the linear growth rate of channel flows and shows close to third-order convergence in agreement with results obtained with the static grid code {\small ATHENA} (see \cref{fig:linearGrowthL1}).
The Lagrangian method MFM on the other hand requires for similarly accurate results a much higher number of resolution elements and also the relatively large Wendland C4 kernel with 200 neighbours, which implies much higher computational costs.

After the exponential growth in the linear regime, the MRI saturates, and provided it does not die out a quasi-stationary turbulent state forms  that can be described by time-averaged quantities. The exact behaviour depends on the numerical resistivity of the code, which can be increased/decreased in our simulations by imposing a stronger/weaker divergence cleaning. Stronger numerical resistivity leads typically to a weaker MRI, causing smaller average magnetic and turbulent kinetic energies as well as smaller stresses, and therefore weaker angular momentum transport. A stronger cleaning on the other hand also means that errors due to deviations from the condition $\nabla \cdot \bm B = 0$ are smaller.

In unstratified simulations with a net vertical field (NF), the MRI always survives even with strong cleaning and a rather small resolution of 16 resolution elements per scale height. This is in contrast to unstratified simulations without net field (ZNF), in which the MRI can die out for strong cleaning, especially for small boxes. However, a higher resolution can help in this case to sustain the MRI and increase its strength. This is in contrast to previous results for static grid codes, which found a non-convergence \citep{fromang2007mhd} of the saturated quantities of the MRI with increasing resolution. The authors explain this by a decreasing numerical Prandtl number with resolution, which also influences the strength of the MRI. Our results on the other hand show in this respect a more similar behavior to results obtained with SPH \citep{wissing2021magnetorotational}, which hints that the numerical Prandtl number might scale differently between static and moving mesh codes. 

We also performed unstratified ZNF simulations with larger boxes in vertical direction similar to \cite{shi2016saturation}. We find in this case a large-scale mean-field dynamo similar to the results obtained with {\small ATHENA}, and in contrast to the SPH results in \cite{wissing2021magnetorotational}. We attribute this to an active shear current effect, since we find for the transport coefficient $\eta_{yx} <0$, in contrast to \cite{wissing2021magnetorotational} who found $\eta_{yx} >0$. This dynamo increases the strength of the MRI, and only for our lowest resolution the MRI can die out for strong cleaning.

We furthermore carried out stratified shearing box simulations that include the vertical component of the gravitational force of the central object. Due to their higher computational cost we only performed 4 simulations with different resolutions and cleaning strengths.
In all simulations, we find an active $\alpha \omega $ dynamo with a time-varying mean field in the mid-plane. The sign of this mean field changes with a period of around 15 orbits and leads to the characteristic butterfly diagram in the space-time diagram of the mean azimuthal field (see \cref{fig:temporal_avg_stratified}). In all our simulations the turbulence survived for at least 200 orbits (the time we stopped the simulations), and we find a magnetically dominated corona and an MRI-dominated mid-plane in agreement with previous results from the literature. Curiously, we find a bump in the magnetic energy in the boundary region. The MRI is in general a bit weaker in our runs in comparison to previous results for grid codes, and our results are in somewhat closer correspondence to simulations  with strong cleaning in SPH \citep{wissing2021magnetorotational}. We also analyzed the vertical profiles of the magnetic transport coefficients and found good agreement with previous results in the literature, and again we find an an effective shear current effect, in contrast to \citet{wissing2021magnetorotational}.

\begin{figure*}
    \centering
    \includegraphics[width=0.9\linewidth]{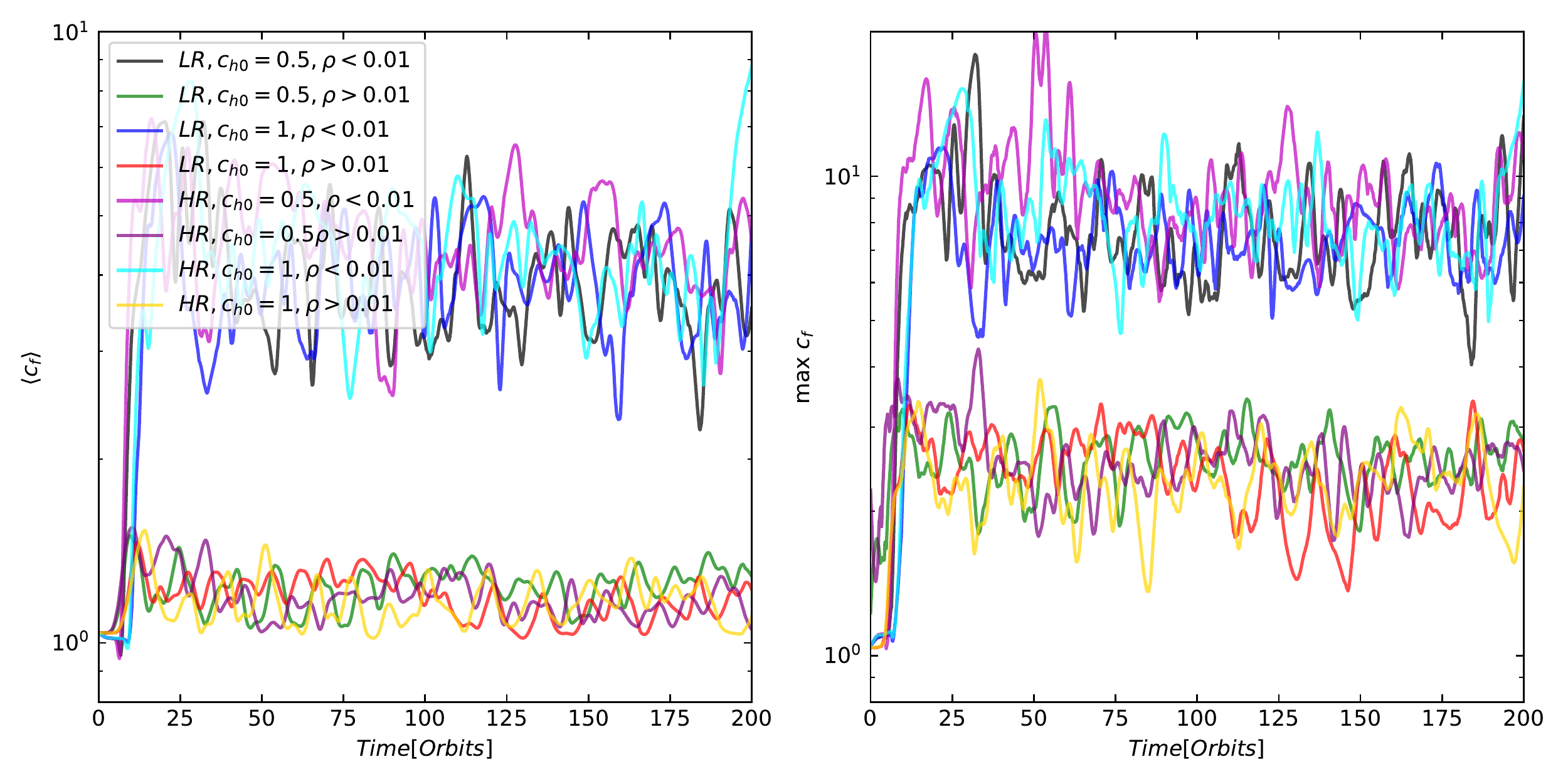}
    \caption{Temporal evolution of the volume-averaged (left) and maximum (right) signal speed (see equation~\ref{eq:signalSpeed}) for our simulations in a stratified shearing box. We split the gas into a low-density component and a high-density component that resides close to the mid-plane. The maximum signal speed is a factor of around 3 larger in the low-density region compared to the high-density region, and around 8 times higher than the average signal speed in the high-density regions. Since the former sets the cleaning speed of the Dedner cleaning, our scheme applies a much stronger cleaning in the high-density region than is in principle required locally. For visual clarity, we have smoothed the curves over 4 orbits using a Savitzky–Golay filter.}
    \label{fig:signalSpeedEvolution}
\end{figure*}

To stabilize our code against divergence errors, we use the Dedner cleaning approach with a globally constant cleaning speed equal to the largest signal speed in the simulation. While in unstratified simulations the box is on average homogeneous and the differences in the signal speeds are smaller, this changes in stratified simulations. Here the corona is magnetically dominated and therefore the maximum signal speed is much larger than the typical sound speed (see \cref{fig:signalSpeedEvolution}). On the one hand, this leads to a stronger than required cleaning in the mid-plane, and therefore larger numerical resistivity and a weaker MRI. On the other hand, this also leads to smaller time steps in the mid-plane, which makes such simulations more expensive. 
We therefore plan to implement alternative cleaning methods in  future work  \citep{tricco2016constrained,hopkins2016constrained} that only require a local cleaning speed in the moving-mesh case. We also note that for global disk simulations we can revert to the standard Powell cleaning in {\small AREPO}, which does not suffer from those disadvantages. 

Especially when we compare the resolution of our simulations with the one from static grid codes, we have to note that for our moving-mesh code the computational costs per resolution element are much larger for pure MHD simulations.
Additional time has to be spent on the construction of the Voronoi mesh and on average cells have more faces than in a Cartesian grid which means there are more Riemann problems to solve.
The unstructured mesh requires the more expensive higher order flux integration introduced in \cite{zier2022simulating} and also reduces the efficiency of memory accesses.
Particle methods suffer from similar overhead and we, therefore, expect similar performance.
As \cite{deng2019local} already noted the performance differences are highly problem dependent and if additional physical effects such as self-gravity dominate the total computational costs, the moving-mesh method will become more competitive in comparison to static grid codes.

Our main results can be summarized as follows:
\begin{itemize}
    \item We find close to third order convergence for linear growth rates of channel modes with absolute errors almost identical to results obtained with  {\small ATHENA}.
    \item The strength of the saturated state of the MRI decreases with stronger numerical resistivity (larger cleaning speed) but deviations from $\nabla \cdot \bm B$ also decrease.
    \item In unstratified NF simulations the MRI does not die out even for strong cleaning and low resolution.
    \item In small, unstratified ZNF simulations the MRI can die out for strong cleaning. The strength of the MRI however increases with higher resolution in contrast to results from static grid codes.
    \item In large, unstratified ZNF simulations we find a large-scale mean-field dynamo \citep[in agreement with][]{shi2016saturation} and an active shear current effect.  The former is significantly weaker in SPH simulations presented by \cite{wissing2021magnetorotational}, which could be caused by a missing shear current effect in those simulations.
    \item We find the characteristic butterfly diagram in stratified simulations and can sustain turbulence for at least 200 orbits. The qualitative results compare well with previous results in the literature, though our MRI is a bit weaker. We attribute this to a too strong cleaning in the mid-plane due to a globally constant Dedner cleaning speed.
\end{itemize}

All in all, our results confirm the high accuracy of our moving-mesh approach for demanding simulations such as MRI-driven turbulence in accretion disks. Our results show reassuring consistency with mesh-based findings, although the relatively high numerical resistivity of the Dedner cleaning approach compared to constrained transport approaches shows up in some of our results. However, our method is readily applicable and well adjusted to global disk simulations, where it represents a very competitive alternative to Eulerian mesh codes. Unlike in the shearing box case, in such simulations we can furthermore employ the Powell scheme for divergence control in our code, which is significantly less diffusive. It thus appears promising to consider full accretion disk calculations that account for the MRI with  {\small AREPO} in future work.

\section*{Acknowledgements}

The authors acknowledge helpful discussions with R\"udiger Pakmor.
We thank the anonymous referee for insightful and constructive comments that helped to improve the paper.

\section*{Data Availability}
The data underlying this paper will be shared upon reasonable request to the corresponding author.

\begin{appendix}

\renewcommand{\thefigure}{A\arabic{figure}}
\setcounter{figure}{0}

\begin{figure*}
    \centering
    \includegraphics[width=0.9\linewidth]{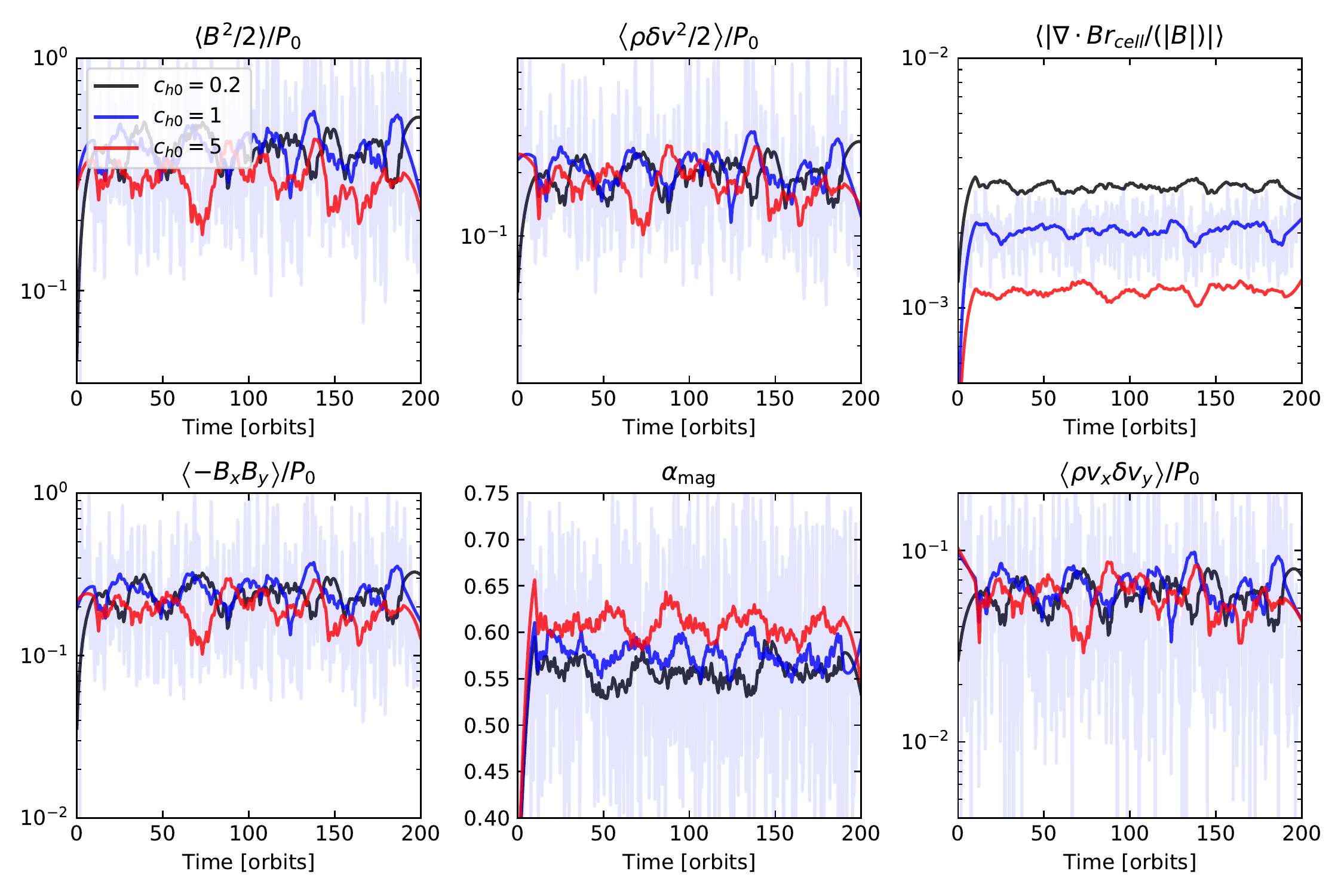}
    \caption{The temporal evolution of several volume weighted quantities for unstratified simulations with a net vertical magnetic field, for a box size $1\times 4 \times 1$ and $48$ cells per scale height. We vary the strength of the Dedner cleaning $c_{h0}$, as indicated in the legend.  The shown quantities are (from left to right, and top to bottom): Magnetic field energy density, kinetic energy density, relative $\nabla \cdot \bm B$ error, Maxwell stress, normalized Maxwell stress (\ref{eq:defAlphaMag}) and Reynolds stress. We have smoothed the curves over 20 orbits using a Savitzky–Golay filter for visual clarity, and show the original curve for one example as a transparent line.}
    \label{fig:evolution_net_field_small}
\end{figure*}

\begin{figure*}
    \centering
    \includegraphics[width=0.9\linewidth]{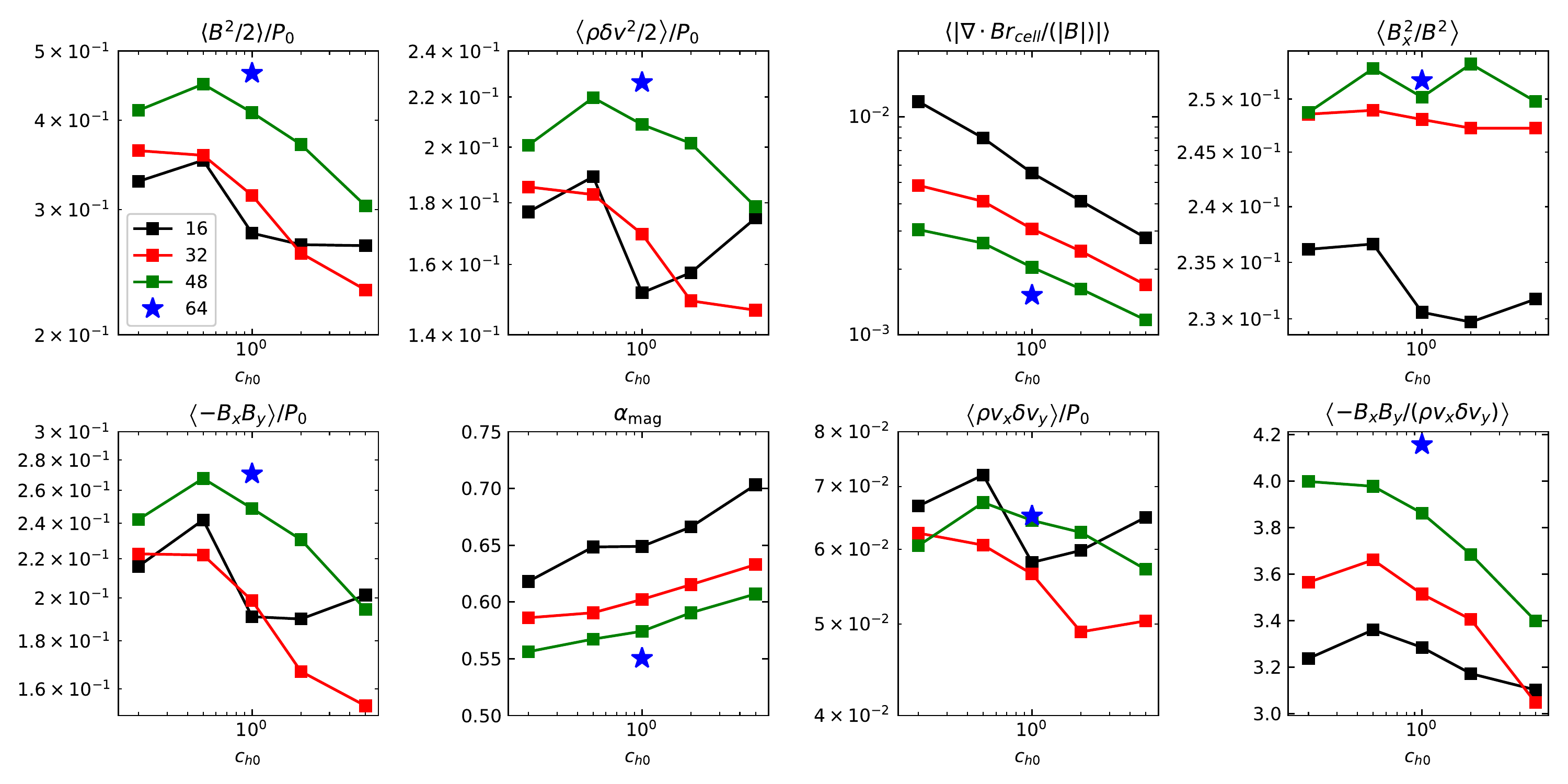}
    \caption{The temporal average of different quantities for unstratified simulations with a net vertical magnetic field and box size $1\times 4 \times 1$, as a function of the strength $c_{h0}$ of the Dedner cleaning. All quantities are averaged over 150 orbits starting after 50 orbits.  We also vary the resolution by using 16, 32, 48 and 64 cells per scale height, as indicated in the legend. For the highest resolution we only performed one simulation with $c_{h0} = 1$ due to the high computational cost of the corresponding run. Besides the quantities already displayed in \cref{fig:evolution_net_field_small}, we also give the ratio between radial and total magnetic field energy (top right), and the ratio of the Maxwell and Reynolds stress (bottom right).}
    \label{fig:overview_avg_NetFlux_small}
\end{figure*}

\bibliographystyle{mnras}
\bibliography{main.bib}

\begin{figure}
    \centering
    \includegraphics[width=1\linewidth]{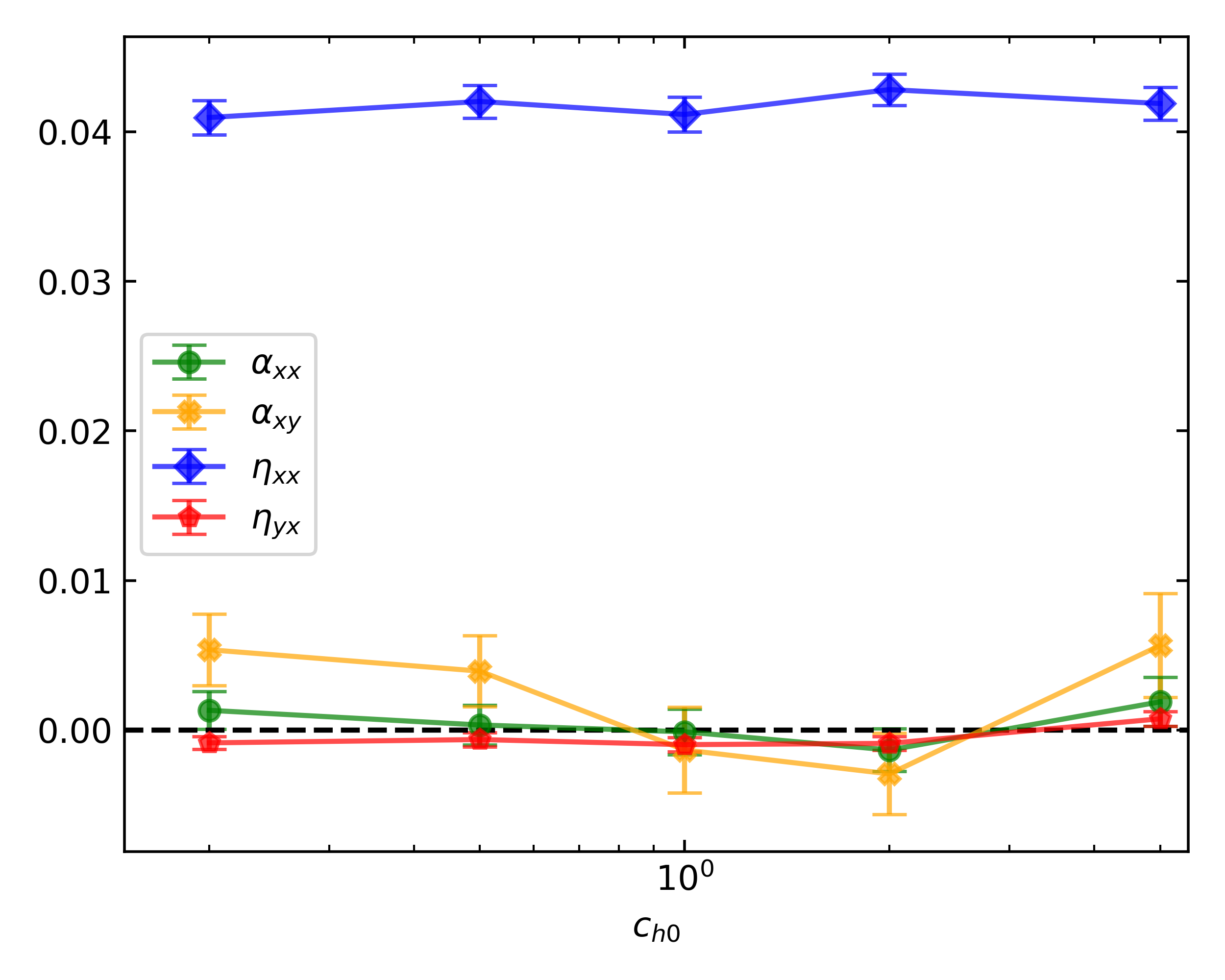}
    \caption{Temporally and spatially averaged transport coefficients as a function of the cleaning strength $c_{h0}$ for unstratified simulations with background field, 48 cell per scale height resolution, and a box size $L_x \times L_y \times L_z = 1 \times 4 \times 1$. The coefficients are averaged over a period of 150 orbits starting after 50 orbits. We additionally included the statistical error of the mean value for each coefficient. As expected, $\eta_{xx}$ is the only coefficient differing significantly from zero.}
    \label{fig:smallBoxNetFluxTranposrtCoeff}
\end{figure}

\section{Net flux MRI in unstratified, smaller box}
\label{app:netFluxMRIInSmallerBox}

In this appendix, we show additional results for unstratified simulations with box size $L_x \times L_y \times L_z = 1 \times 4  \times 1 $, a standard resolution of $16 \times 64 \times 16$ cells, and initial plasma beta $\beta = 330$. In \cref{fig:evolution_net_field_small}, we show the temporal evolution for three different cleaning strengths $c_{h0}$ and a resolution of 48 cells per scale height. In \cref{fig:overview_avg_NetFlux_small}, we show different  spatially and temporally averaged properties describing the MRI as a function of $c_{h0}$ for different resolutions. \cref{fig:smallBoxNetFluxTranposrtCoeff} shows the measured transport coefficients as a function of the cleaning strength. As in the larger box all coefficients except $\eta_{xx}$ are compatible with zero.

\end{appendix}

\bsp	
\label{lastpage}
\end{document}